\documentclass[sigconf]{acmart}
\makeatletter                   %1
\def\mdseries@tt{m}             %1
\makeatother                    %1

\usepackage{array}
\usepackage{xspace}
\usepackage{booktabs} % For formal tables
\usepackage{listings}
\usepackage{mathtools}
\usepackage{subfig}
\usepackage[T1]{fontenc}
\usepackage{multirow}
\usepackage{hyperref}
\usepackage{enumitem}
\usepackage{graphicx}
\usepackage{dblfloatfix}

\usepackage[frozencache]{minted}
\fvset{linenos, fontsize=\small, frame=lines, numbersep=1pt, xleftmargin=.75em}
\DeclareCaptionSubType{listing} % Allow sublistings

\renewcommand\footnotetextcopyrightpermission[1]{} % removes footnote with conference information in first column
\setcopyright{rightsretained}
\newcommand{\figref}[1]{Fig.~\ref{fig:#1}}
\newcommand{\tabref}[1]{Tab.~\ref{tab:#1}}
\newcommand{\lstref}[1]{Lst.~\ref{lst:#1}}

\newcommand{\secref}[1]{Sec.~\ref{sec:#1}}

\usepackage[per-mode=symbol]{siunitx}
\DeclareSIUnit \bit {bit}
\DeclareSIUnit \bits {bits}
\DeclareSIUnit \byte {B}
\DeclareSIUnit \Byte {Byte}
\DeclareSIUnit \Bytes {Bytes}
\DeclareSIUnit \cycle {cycle}
\DeclareSIUnit \cycles {cycles}
\DeclareSIUnit \hz {Hz}
\DeclareSIUnit \op {Op}
\DeclareSIUnit \flop {FLOP}
\DeclareSIUnit \operand {operand}
\DeclareSIUnit \operands {operands}
\DeclareSIUnit \transfer {T}
\DeclareSIUnit \cell {cell}

\usepackage{soul}
\definecolor{lightyellow}{RGB}{250, 250, 180}
\sethlcolor{lightyellow}
\soulregister\cite7
\soulregister\ref7
\soulregister\pageref7
\soulregister\figref7
\soulregister\secref7
\soulregister\tabref7
\soulregister\lstref7
\soulregister\ckr7
\soulregister\cks7

\newcommand{\cameraready}[1]{#1}
\newcommand{\smipop}{\texttt{SMI\_Pop}\xspace}
\newcommand{\smipush}{\texttt{SMI\_Push}\xspace}

\def\maybespacelist{.,;!?}
\def\maybespace{\let\nexxt=\space 
	\edef\maybespaceA{\noexpand\maybespaceB\maybespacelist\relax}%
	\futurelet\next\maybespaceA
}
\def\maybespaceB#1{\ifx#1\relax \nexxt \else \ifx#1\next \let\nexxt=\relax\fi
	\expandafter\maybespaceB\fi
}
\protected\def\cks{\textsf{CK\textsubscript{S}}}
\protected\def\ckr{\textsf{CK\textsubscript{R}}}
\newcommand{\pop}{\texttt{Pop}\xspace}
\newcommand{\push}{\texttt{Push}\xspace}

\copyrightyear{2019}
\acmYear{2019}
\acmConference[SC '19]{The International Conference for High Performance Computing,
Networking, Storage, and Analysis}{November 17--22, 2019}{Denver, CO, USA}
\acmBooktitle{The International Conference for High Performance Computing, Networking,
	Storage, and Analysis (SC '19), November 17--22, 2019, Denver, CO, USA}
\acmPrice{15.00}
\acmDOI{10.1145/3295500.3356201}
\acmISBN{978-1-4503-6229-0/19/11}

\begin{document}
  \title[Streaming Message Interface: Distributed Programming on Reconfigurable Hardware ]{Streaming Message Interface: High-Performance Distributed Memory
  Programming on Reconfigurable Hardware}
  
  \author{Tiziano De Matteis}
  %\authornote{Dr.~Trovato insisted his name be first.}
  \affiliation{%
    \institution{Department of Computer Science, ETH Zurich}
    %\city{ETH Zurich}
    %\state{Switzerland}
  }
  \email{tiziano.dematteis@inf.ethz.ch}
  
  \author{Johannes de Fine Licht}
  \affiliation{%
    \institution{Department of Computer Science, ETH Zurich}
  }
  \email{definelicht@inf.ethz.ch}

  \author{Jakub Ber\'anek}
  \affiliation{%
    \institution{IT4Innovations, V\v{S}B - Technical University of Ostrava}
  }
  \email{jakub.beranek@vsb.cz}
  
  \author{Torsten Hoefler}
  \affiliation{%
    \institution{Department of Computer Science, ETH Zurich}
  }
  \email{htor@inf.ethz.ch}
  \renewcommand{\shortauthors}{T. De Matteis et al.}
  
  \hyphenation{pi-pe-lin-ing}
  \keywords{Distributed Memory Programming, Reconfigurable computing, High-Level Synthesis Tools}
  
% [It must be <= 150 words] 
% Network-enabled FPGAs are becoming common in data-centers and supercomputers.
% Inspired by classical message-passing model, we introduce a Streaming
% Messages, as a programming model for communicating application a  Interface
% (SMI), a communication model and a small set of communication primitives for
% enabling distributed memory programming on FPGAs and, more in general,
% reconfigurable hardware.  We introduce a reference implementation realized
% with high-level synthesis tools.

\begin{abstract}
  Distributed memory programming is the established paradigm used in 
  high-performance computing (HPC) systems, requiring explicit communication
  between nodes and devices.
  When FPGAs are deployed in distributed settings, communication is typically
  handled either by going through the host machine, sacrificing performance, or
  by streaming across fixed device-to-device connections, sacrificing
  flexibility.
  We present Streaming~Message~Interface~(SMI), a communication model and API
  that unifies explicit message passing with a hardware-oriented programming
  model, facilitating minimal-overhead, flexible, and productive inter-FPGA
  communication.
  Instead of bulk transmission, messages are streamed across the network during
  computation, allowing communication to be seamlessly integrated into pipelined
  designs.
  We present a high-level synthesis implementation of SMI targeting a dedicated
  FPGA interconnect, exposing runtime-configurable routing with support for
  arbitrary network topologies, and implement a set of distributed memory
  benchmarks. 
  Using SMI, programmers can implement distributed, scalable HPC programs on
  reconfigurable hardware, without deviating from best practices for hardware
  design.  
  %SMI utilizes direct device-to-device connections, while adding a
  %runtime-configurable communication layer that enables scalability, and
  %supports arbitrary network topologies.
\end{abstract}
  
\maketitle

\begin{figure}[t]
  \includegraphics[width=\columnwidth]{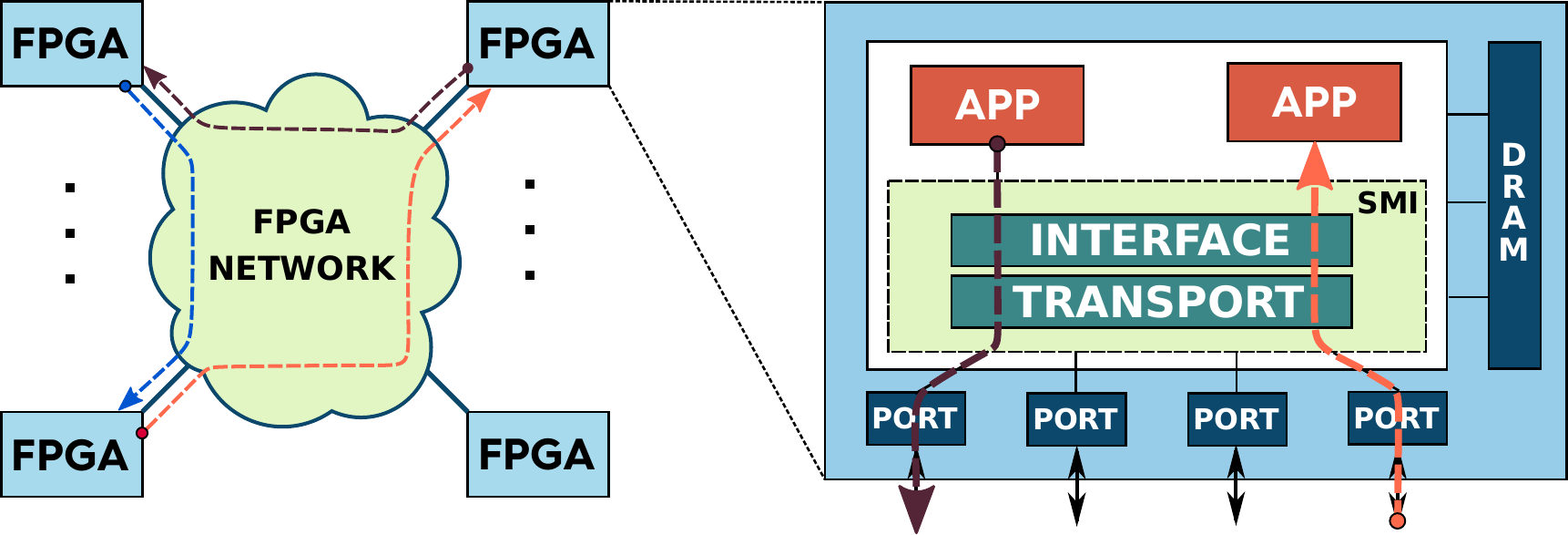}
  \label{figure:smi}
  \vspace{-1em}
  \caption{Multi-FPGA programming with SMI\protect\footnotemark.}
  %\vspace{-1em}
\end{figure}

% Putting the figure all the way up is there only way latex is convinced to put
% it on the second page (gosh)
\begin{figure*}[!b]
  \begin{minipage}[b]{.33\textwidth}
    \begin{minted}[linenos=False]{C++}
for (int i = 0; i < N; i++)
  buffer[i] = compute(data[i]);
SendMessage(buffer, N, my_rank + 2);
    \end{minted}
    \vspace{0.5em}
    \centering
    \includegraphics[width=\textwidth]{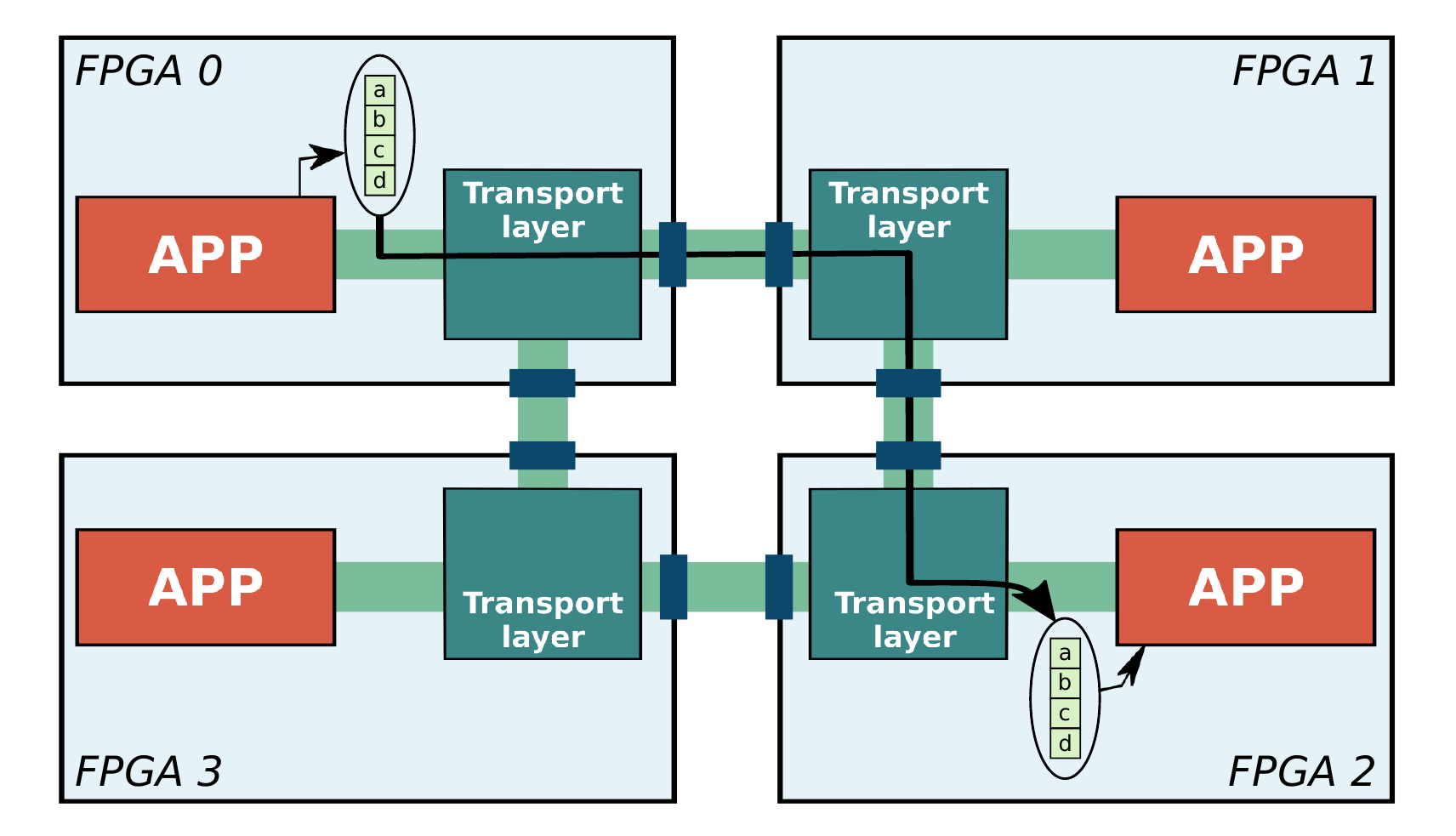}
    \vspace{-2.5em}
    \caption{Message passing.}
    \label{fig:message_passing}
  \end{minipage}\hfill%
  \begin{minipage}[b]{.33\textwidth}
    \begin{minted}[stripnl=false, linenos=False]{C++}
// Channel fixed in the architecture
for (int i = 0; i < N; i++)
  stream.Push(compute(data[i])); 
    \end{minted}
    \vspace{0.5em}
    \centering
    \includegraphics[width=\textwidth]{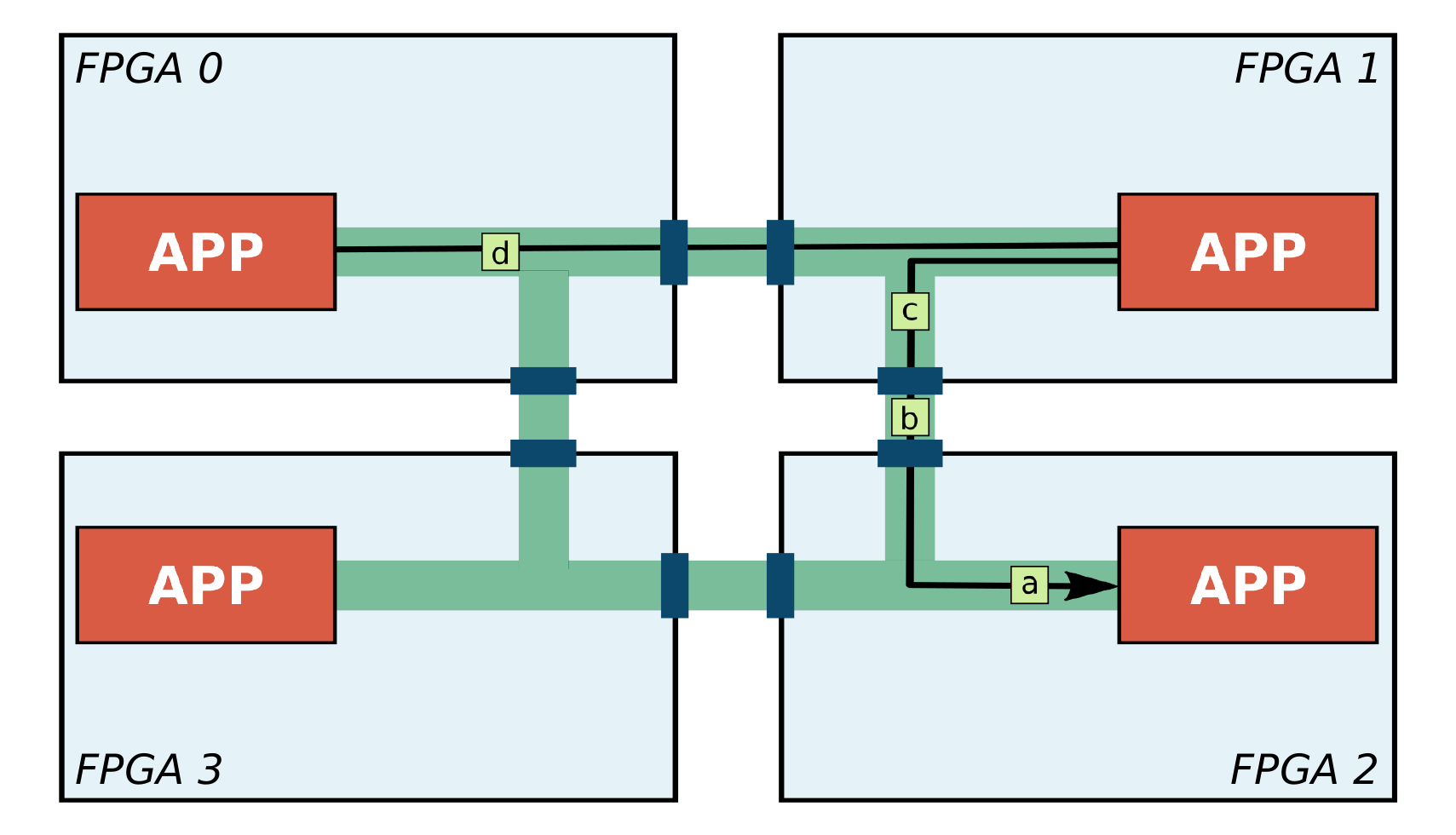}
    \vspace{-2.5em}
    \caption{Streaming.}
    \label{fig:streaming}
  \end{minipage}\hfill%
  \begin{minipage}[b]{.33\textwidth}
    \centering
    \begin{minted}[linenos=False]{C++}
Channel channel(N, my_rank + 2, 0);
for (int i = 0; i < N; i++)
  channel.Push(compute(data[i])); 
    \end{minted}
    \vspace{0.5em}
    \centering
    \includegraphics[width=\textwidth]{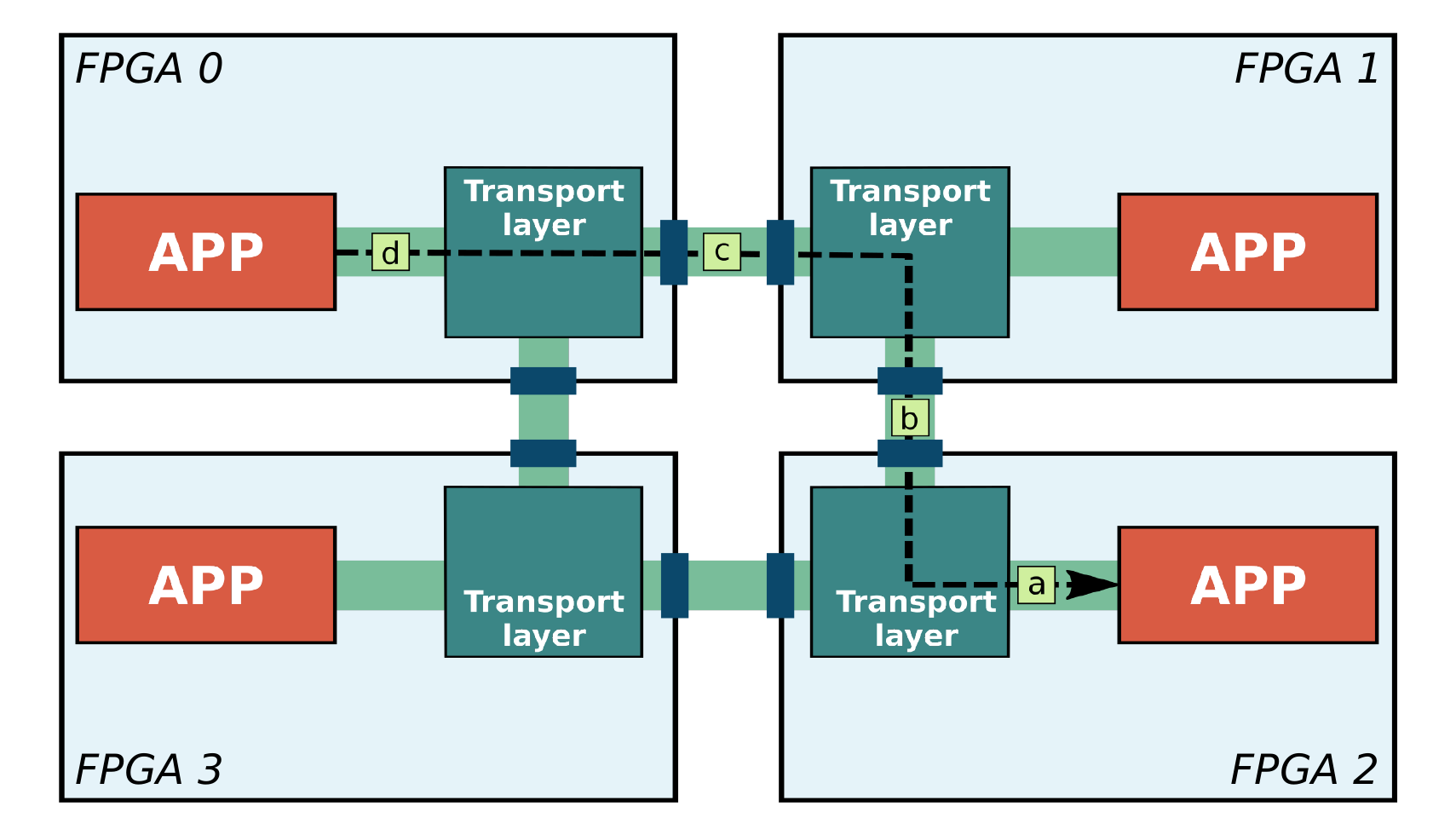}
    \vspace{-2.5em}
    \caption{Streaming messages.}
    \label{fig:streaming_messages}
  \end{minipage}
\end{figure*}
  
\section{Introduction}

%\para{reconfigurable/programmable hardware is important}
The end of Moore's law and Dennard scaling causes a major disruption to the
high-performance computing industry. Both require us to re-think computer
architecture in order to reduce data movement and power dissipation on chips,
and to use the existing transistors more efficiently.
To address both problems, reconfigurable architectures with application-specific
dataflow as well as compute logic provide a viable option.
Many large-scale datacenter operators, such as Amazon~\cite{amazon_f1} and
Microsoft~\cite{brainwave}, already build on reconfigurable logic to specialize
hardware implementations to their workloads.
Reconfigurable logic can avoid the well-known architectural von~Neumann
(load-store) bottleneck, but poses new challenges in programming these devices. 
Yet, highly efficient designs with significant performance and energy benefits
have shown that efforts in the area are well spent~\cite{kara2017fpga,
cnn_cluster_fpga, finn}.

\footnotetext{SMI is publicly available at \url{ https://github.com/spcl/SMI}\\ This is a preprint version.}

%\para{we focus on HLS on FPGAs}
Reconfigurable hardware traditionally came in the flavor of fully
configurable logic devices, field-programmable gate arrays (FPGAs), or
devices with a large fraction of hardened logic with flexible datapaths,
so-called coarse-grained reconfigurable arrays (CGRAs).
Today, the boundary between FPGAs and CGRAs is vanishing, with the introduction
of high-performance FPGAs that contain powerful DSP cores, such as Intel's
Stratix~10 ($\SI{10}{\tera\flop\per\second}$ single
precision~\cite{stratix10_product_table}), and AI~Engines in Xilinx Versal
devices~\cite{versal_product_table}.
% and Xilinx' Kintex UltraScale (8.2 Tflop/s single precision~\cite{}).
% ???  https://www.xilinx.com/support/documentation/white_papers/wp434-ultrascale-smarter-systems.pdf (page 4???)
Those high-performance FPGAs become highly attractive for HPC workloads
if the programming problem can be solved. 
Recent developments in high-level synthesis promises to deliver high
productivity on FPGAs replacing the traditional register transfer logic (RTL)
specification with a standard C/C++ code interface integrated with modern OpenCL
frameworks~\cite{intel_opencl, xilinx_hls}. 

%\para{most FPGAs have dedicated network connections but are not
%programmable in a portable way}
Most of the HLS research focuses on programming one or multiple FPGAs
attached to a single host.
Yet, in HPC systems, single FPGAs need to be scaled up to clusters
containing many devices.
Today, communication is performed through message passing at the host, where
data is usually transported via PCI Express (PCIe) to the main memory, and then
through a different PCIe channel to the network interface. This adds high
overheads in terms of latency, bandwidth, and load on the host's memory
subsystem.
Nearly all modern FPGA chips bear high-performance serial link network
connections. For example, Intel's Stratix~10 chip has four
$\SI{40}{\giga\bit\per\second}$ connections and Xilinx' UltraScale+ chips
support $\SI{30}{\giga\bit\per\second}$ off-chip connectivity. 
These links are often available via proprietary interfaces for communication
among directly connected FPGAs. 
Unfortunately, no distributed memory programming model exists for HLS-programmed
devices, and programmers are forced to resort to licensed IP cores and RTL
designs to implement FPGA-to-FPGA communications~\cite{inference_tree,
opecl_ready_network}.

%\para{SMI to the rescue}
We propose a distributed memory HLS programming model for FPGAs that provides
the convenience of message passing for HLS-programmed hardware devices. 
While we cannot simply use the Message Passing Interface (MPI) API due to the
peculiarities of programmed hardware, we are heavily inspired by MPI's
interface, to benefit from its proven effectiveness in practice, and maintain
familiarity for programmers.
The reason for this specialization is that high-performance HLS designs are
deeply pipelined and vectorized. This means that several results are produced at
each clock cycle and shallow buffering along predefined (pipeline) paths is a
necessity for performance.
Thus, our \emph{Streaming~Message~Interface} (SMI) does not assume that buffers are
first computed and then communicated---instead, sending a message is
integrated into the pipeline.
The key concept of SMI is its streaming nature, where a send or receive is
set up first, and the data is then written or read on a cycle-by-cycle basis.
This concept modifies MPI-style messages into transient channels, that have
similar semantics, but integrate seamlessly with HLS-programmed pipelines.

% \para{we're all great - maybe redundant, need to see}
% - we define the conceptual interface of SMI for FPGAs that integrates
% well with OpenCL and other HLS approaches
% - we implement and test the interface with microbenchmarks and
% application kernels on a real Stratix 10 FPGA cluster
% - we see speedups of XXX\% over standard host-based communication (need
% to measure)
% - we release our implementation as open source for the benefit of the
% community
% - we really need a poster-child figure ... maybe performance, maybe
% overview?

\noindent The key contributions of our work are:
\begin{itemize}[leftmargin=*]
  \item We propose the \emph{streaming messages} communication model, unifying
  the message passing and streaming models;
  \item We design the Streaming Message Interface (SMI), an HLS communication
  interface specification for programming streaming messages in distributed
  memory multi-FPGA systems;
  \item We implement and benchmark a reference implementation of SMI that
  integrates with OpenCL on Intel FPGAs;
  % \item We develop concepts for FPGA-based routing to support switch-less
  % operation; \tiziano{We should highlight this more in the routing section}
  %\item We show speedups over OpenCL for microbenchmarks and application  kernels; %\htor{quantify!} \tiziano{we can remove this? We can only mention improvements on bandwidth and latency}
  \item We release the reference library and example applications implemented
  with a modern HLS tool as open source code that does not rely on additional
  licensed IP cores.
\end{itemize}
We evaluate our approach on several numerical computations, showing the
performance benefits of distributed memory FPGA programming, by increasing
available compute resources and memory bandwidth. 
  
% \paragraph{Paper structure.}
% We can describe the library as organized in a sort of "network stack": at the
% top-most level we have an \textit{high-level interface} layer, which offers an
% API that satisfies our programming model. Below it, we find a \textit{transport}
% layer, a hardware library which actually implements the communication
% primitives. Then, at the lower-most level we have the \textit{physical} layer.
% In our case, the physical layer is what the chip and BSP offers (a channel abstraction for QSFP connectors). The top-most layer is \textit{software}, while the other two are \textit{hardware} layers.
%
% This organization could also be reflected in the sections of the core part of the paper.

\section{Programming FPGA Communication}

To design a suitable communication model for distributed FPGA programming, we
wish to learn from the most prominent model found in HPC, namely message passing
(specifically, MPI), but adapt it to a form suitable for hardware programming.
We call our model \emph{streaming messages}, and will introduce it by
highlighting the gaps in existing models, which it has been designed to fill.
%
% \subsection{FPGA programming model}
% \label{sec:programming_fpgas}

Programming FPGAs with high-level synthesis revolves around designing deep
hardware \emph{pipelines}, exploiting the spatially parallel nature of the FPGA
fabric. Parallelism is achieved by making this pipeline deeper (pipeline
parallelism), by making the pipeline wider (vector parallelism), or by
replicating the pipeline entirely (task parallelism)~\cite{hls_transformations}.
Pipelines are expressed as loops in the HLS code, designed such that new
operands can be accepted every cycle. It is thus imperative that a communication
model is compatible with this programming model, allowing communication to
happen during pipelined computations.

\subsection{Existing Communication Models}
\subsubsection{Message Passing}
\label{sec:message_passing}

The paradigm of \emph{message passing} uses local buffers to both send and
receive information to/from other processes (called \emph{ranks}). A distributed
algorithm will work on a local subset of data on each rank, then indicate to the
communication layer when a buffer is ready to be sent to another rank, or when
it is ready to receive new data into a buffer. This is illustrated with an
example code in \figref{message_passing}, where a buffer is populated in a loop,
then sent to another rank. To hide communication time, message passing uses
\emph{non-blocking} calls to overlap communication and computation, thus letting
ranks operate on different data than what is currently being exchanged.

In the context of hardware programming, message passing has two key
shortcomings. First, the model relies on \emph{bulk} transfers, which is a poor
match to the HLS programming model, as we wish to communicate during pipelined
computation. Second, bulk transfers imply large buffers required to store
intermediate data. On the CPU, these buffers exist in the global memory space,
and can dynamically move between cache and DRAM, depending on their size and the
behavior of the program. In contrast, buffers used when programming for hardware
are explicitly instantiated in a fast memory distributed across the chip, and
moving them to an off-chip memory requires explicit wiring to limited DRAM
interfaces, which are shared among all accesses. Fully adapting this approach in
hardware would thus come with significant disadvantages in resource utilization,
programmability, and performance.

\subsubsection{Streaming}
\label{sec:streaming}

A classical way of moving data between FPGAs is to simply \emph{stream} it
across an inter-FPGA channel in a pipelined fashion (e.g., the Maxeler dataflow
engine architecture~\cite{maxeler}), similar to how data is moved across the
chip on a single FPGA. This approach offers a way of expressing communication
that is natural to the hardware paradigm, by pushing data to the output
interface in a pipelined fashion during processing (see~\figref{streaming}).
Streaming relies on point-to-point connections known at configuration time,
suitable for extending one-dimensional \emph{systolic~array}-style architectures
across multiple chips~\cite{sano2014multi}.

The major shortcoming of streaming interfaces in a distributed memory setting is
the lack of flexibility in the implied API and transport layer. Even if the
target platform has the necessary hardware ports, a shell that exposes them, and
an API to access them, moving data from a given source to a given destination
requires the programmer to construct the \emph{exact path} that the data has to
move across as part of the architecture. This has to be repeated for every
desired communication channel, for every target application; including
forwarding logic when multiple hops are required, and arbitration between
different channels using the same hardware connection.
In the example shown in \figref{streaming}, data travelling from
FPGA~0 to FPGA~2 must first be sent through FPGA~1, where custom user logic must
take care of forwarding it to the final destination.
For more complicated distributed memory environments, the streaming interface in
its pure form is thus insufficient to productively express arbitrary 
communication patterns.

\subsection{Streaming Messages}
\label{sec:streaming_messages}

To capture the key ideas of message passing and streaming, while addressing the
gaps in both outlined above, we introduce \emph{streaming messages}: an
HPC-oriented communication model for hardware programming, with an implied
transport layer.  Streaming messages replace traditional, buffered messages with
pipeline-friendly \emph{transient channels}, offering a streaming interface to
the hardware programmer, but with the flexibility known from the message passing
paradigm. Knowledge of the interconnect topology is not required at
compile-time: channels between endpoints are transiently established, where
source and destination ranks can be specified dynamically.  This is illustrated
in \figref{streaming_messages}, where a kernel on rank~0 on $\text{FPGA}_0$
opens a channel to rank~2 on $\text{FPGA}_2$, using port~0 to distinguish the
target application (akin to starting a non-blocking send in MPI, but without
implying that the data is ready), then pushes data to the channel during
processing in a pipelined fashion (as in the streaming paradigm).  Routing data
to the destination is then handled transparently by the transport layer.

In streaming messages, a \emph{rank} is associated with a coarse
hardware entity assigned to dedicated communication logic, connected to the
incoming and outgoing hardware communication ports. A \emph{port} uniquely
identifies an endpoint \emph{within} a rank, and implements a hardware streaming
interface for every \texttt{Push} and \texttt{Pop} operation present in the code
to/from a matching external port. This implies that all ports must be known at
compile time, such that, within each rank, the necessary hardware connections
between the communication endpoints and the network can be instantiated.  Ports
must be specified both for point-to-point and collective communication
primitives to establish the required hardware.  All ports represent hardware
connections, and can thus operate fully in parallel.
% \johannes{In MPI terms, ports conceptually resemble \emph{tags} for
% point-to-point communication, but differentiate themselves by being
% \emph{static} (as they imply physically instantiated hardware), and by also
% being relevant for collective operations.}

%Streaming messages use an adapted version of MPI's rank and tag terminology. A \emph{rank} is associated with a coarse hardware entity assigned to dedicated communication logic, connected to the incoming and outgoing hardware communication ports.
%A \emph{tag} represents an endpoint \emph{within} a rank, and implements a hardware streaming interface for every \texttt{Push} and \texttt{Pop} operation present in the code to/from a matching external tag. This implies that all tags must be known at compile time, such that, within each rank, the necessary hardware connections between the communication endpoints and the network can be instantiated. 
Channels can be programmed either in a \emph{single program, multiple data}
(SPMD) fashion, or in a \emph{multiple program, multiple data} (MPMD) fashion.
In this work, we assume a single rank per FPGA. Ranks involved in communication
and the total number of ranks can then be dynamically altered without
recompiling the program, by simply updating the routing configuration at each
rank.

\section{Streaming Message Interface}
\label{sec:smi}

To concretize the concept of streaming messages, we introduce the Streaming
Message Interface (SMI), a communication interface specification for HLS
programs inspired by MPI~\cite{mpi_standard}.
SMI is not an implementation, and merely implies the functionality that must be
supported by the transport layer to support the interface specification.
The interface exposes primitives for both point-to-point and collective
communications.

\begin{listing}[b]
  \begin{minted}[linenos=False]{C++}
void Rank0(const int N, /* ...args... */) {
  SMI_Channel chs = SMI_Open_send_channel( // Send to
      N, SMI_INT, 1, 0, SMI_COMM_WORLD);   // rank 1
  #pragma ii 1 // Pipelined loop
  for (int i = 0; i < N; i++) {
    int data = /* create or load interesting data */;
    SMI_Push(&chs, &data);
} }
 
void Rank1(const int N, /* ...args... */) {
  SMI_Channel chr = SMI_Open_recv_channel( // Receive
      N, SMI_INT, 0, 0, SMI_COMM_WORLD);   // from rank 0
#pragma ii 1                               // Pipelined loop
  for (int i = 0; i < N; i++) {
    int data;
    SMI_Pop(&chr, &data);
    // ...do something useful with data... 
} }
  \end{minted}
  \vspace{-1em}
  \caption{MPMD program with two ranks.}
  \label{lst:smi}
 \end{listing}

\subsection{Point-to-Point Communication}

% In the streaming message model, programs can stream messages between each other,
% and across multiple FPGAs.
Point-to-point communication in SMI codes is based on transient channels: when
established, a streaming interface is exposed at the specified port at either
end, allowing data to be streamed across the network using FIFO
semantics, with an optional finite amount of buffer space at each endpoint.  A
streaming message consists of one or more elements with a specified data type.
The communication endpoints are uniquely identified by their \textit{rank}.
Ranks uniquely identify FPGA devices, and ports distinguish distinct
communication endpoints within a rank.
%Analogous to MPI, the communication endpoints are uniquely identified by their \textit{rank} and \textit{tag} parameters. Ranks uniquely identify FPGA devices, while tags distinguish distinct endpoints within a rank.

% This comes a bit out of nowhere
% In addition to this, in SMI the tags are also used to identify the proper
% hardware connections between the application and the transport layer.

The example in \lstref{smi} shows an MPMD application composed of two ranks
implemented with SMI (for code examples, we use the Intel FPGA OpenCL directive
syntax, where pragmas apply to the following scope).  Rank 0 streams a message
of $N$ integer elements to Rank 1 using a \textit{send} channel. Rank 1 opens a
\textit{receive} channel to receive the message, and applies a computation on
each data item.
% ^ the above sentence is weird
Input and output channels are opened before the beginning of the loop, and
messages are received and sent one-by-one during computation. Channels are thus
accessible with a streaming cycle-by-cycle interface: computations can
\texttt{Push} or \texttt{Pop} data to/from a communication channel, one data
element per clock cycle.
 
\subsubsection{Point-to-Point Communication API}

The user can declare a \textit{send} or \textit{receive} channel by specifying
the number of elements to send, the data type of the elements, the source or
destination rank, the port, and the communicator. Once established, channels
exist in code in the form of \textit{channel descriptors}. Channels are
implicitly closed when the specified number of elements have been sent or
received.
% The message size is intended as a hint for the transport layer, to give it a
% better opportunity to optimize packet delivery and arbitrate between different
% channels. \tiziano{ I didn't understand this sentence}
%dunno why minted highlight these two keywords
\begin{minted}[breaklines=True, linenos=False]{C++}
SMI_Channel SMI_Open_send_channel(int count, SMI_Datatype type, int destination, int port, SMI_Comm comm);
SMI_Channel SMI_Open_recv_channel(int count, SMI_Datatype type, int source, int port, SMI_Comm comm);
\end{minted}
Analogously to MPI, \emph{communicators} can be established at
runtime, and allow communication to be further organized into logical groups.  
%Like in MPI, communicators allow communication to be further organized in logical groups. 
Channels can also be used to communicate between two applications
that exist within the same rank using matching ports.
% Currently, SMI supports the equivalent of base C data types, e.g, \texttt{int}
% and \texttt{float}.
To send and receive data elements from within the pipelined HLS code, SMI
provides the \smipush and \smipop primitives:
\begin{minted}[linenos=False, breaklines]{C++}
void SMI_Push(SMI_Channel* chan, void* data);
void SMI_Pop(SMI_Channel* chan, void* data);
\end{minted}
Both functions operate on a channel descriptor from a previously opened channel,
and a pointer either to the data to be sent, or to the target at which to store
the data.
These primitives are blocking, such that \smipush does not return before the
data element has been safely sent to the network, and the sender is free to
modify it, and \smipop returns only after the output buffer contains the newly
received data element.

To respect the streaming message model, \smipush and \smipop must be implemented
in such a way that: \textit{i)} data elements are sent and received in the same
order specified by the user, and \textit{ii)} calling them can be pipelined to a
single clock cycle, such that they can be used in pipelined loops without
impairing the initiation interval.
%
%In general, the user may want to send the message element as soon as possible. For this reason, SMI offers a \texttt{SMI\_Push\_flush} call, that
Additionally, the type specified by the \smipush/\smipop operations must match
the ones defined in the \texttt{Open\_Channel} primitives. With these
primitives, communication is programmed in the same way that data is normally
streamed between intra-FPGA modules.

%Push must be able to accept one data element at each clock cycle, so that it can be easily pipelineable.

%Channel communications act as synchronization points?

\subsection{Collective Communication}
\label{sec:collectives}

Collective communication in MPI is key to develop distributed applications that
can scale to a large number of nodes. In collective operations, all ranks in a
given communicator must be involved in communicating data.  SMI defines the
\texttt{Bcast}, \texttt{Reduce}, \texttt{Scatter}, and \texttt{Gather}
collective operation primitives analogous to their MPI counterparts.
% The used arguments resembles the one employed in the MPI equivalent. 

Each collective operation defined by SMI implies a distinct channel
type, open channel operation, and communication primitive.
\cameraready{The example in \lstref{broadcast} shows an SPMD application in which the root rank broadcasts the locally produced elements to the other ranks in the communicator.}
\begin{listing}[htb]
	\begin{minted}[linenos=False]{C++}
void App(int N, int root, SMI_Comm comm, /* ... */) {
  SMI_BChannel chan = SMI_Open_bcast_channel(
      N, SMI_FLOAT, 0, root, comm);
  int my_rank = SMI_Comm_rank(comm);
  for (int i = 0; i < N; i++) {
    int data;
    if (my_rank == root)
      data = /* create or load interesting data */;
    SMI_Bcast(&chan, &data);
    // ...do something useful with data...
} }
\end{minted}
\vspace{-1em}
\caption{SPMD program with broadcast.}
\label{lst:broadcast}
\end{listing}

\noindent To perform a \texttt{Bcast}, each rank opens a
broadcast-specific channel (\texttt{SMI\_BChannel}), indicating the count and
data type of the message elements, the rank of the root, the
port,~and the communicator:
\begin{minted}[linenos=False, breaklines]{C++}
SMI_BChannel SMI_Open_bcast_channel(
    int count, SMI_Datatype type, int port, int root, SMI_Comm comm);
\end{minted}
To participate in the broadcast operation, each rank will use the associated
primitive (analogous to \smipush and \smipop for \texttt{Send} and
\texttt{Recv}, respectively):
\begin{minted}[linenos=False,breaklines]{C++}
void SMI_Bcast(SMI_BChannel* chan, void* data);
\end{minted}
If the caller is the root, it will push the data towards the other ranks.
Otherwise, the caller will pop data elements from the network. 
\cameraready{
Similarly, to perform a \texttt{Reduce}, the associated channel must be opened,
indicating the reduction operation to perform, such as \texttt{SMI\_ADD},
\texttt{SMI\_MAX}, or \texttt{SMI\_MIN}:}
\begin{minted}[linenos=False, breaklines]{C++}
SMI_RChannel SMI_Open_reduce_channel(int count, SMI_Datatype type, SMI_Op op, int port, int root, SMI_Comm comm);
\end{minted}
Data communication occurs with the primitive:
\begin{minted}[linenos=False, breaklines]{C++}
void SMI_Reduce(SMI_RChannel* chan, void* data_snd, void* data_rcv);
\end{minted}
\cameraready{
Each rank sends its contribution (\texttt{data\_snd}), while the reduced result is produced to the root rank (\texttt{data\_rcv}).

SMI allows multiple collective communications \cameraready{of the same type} to execute in parallel, provided that they use separate ports. %Tiz: I have added a conservative "of the same type" because of the packetization: i) data type must be the same ii) due to our implementation reduce does not packetize and this can cause problems when you use it with something else
We leave out the interfaces for \texttt{Scatter} and \texttt{Gather}, as they
follow the same scheme as presented above.}

\subsection{Buffering and Communication Mode}

SMI channels are characterized by an \textit{asynchronicity degree} $k \geq 0$,
meaning that the sender can run ahead of the receiver by up to $k$ data
elements.  If the sender tries to push the $(k+1)$-th element before an element
is popped by the receiver, the sender will stall.  The concrete implementation
of these buffers can use any form of on-chip memory.  Because of this
asynchronicity, an SMI send is \emph{non-local}: it can be started whether or
not the receiver is ready to receive, but its completion \emph{may} depend on
the receiver, if the message size is larger than $k$.
% FIFO buffers, that are used to actually implement our communication channels,
% provide a certain buffer space (tunable at compile time). 
Correctness of the communication in a distributed setting must be guaranteed by
the user, i.e., ensuring that there are no cyclic dependencies between sends and
receives that allow deadlocks, and that the program will terminate even if the
system provides no buffering.

If the channel asynchronicity degree is bigger or equal than the
message size, we suggest to use an \textit{eager} protocol to transfer data for
efficient point-to-point communication:
elements can be pushed into the network without first performing a handshake
with the receiver, aided by buffers at either endpoint.  This saves costly
round-trip latencies, improving the efficiency of small messages.  Creating a
new channel is thus a zero-overhead operation, as this merely instructs the
transport layer where data should be sent. The network interfaces must be able
to handle stalling and backpressure to safely enable eager communication.
On the other hand, if the buffer size is smaller than the message
size, a transmission protocol with credit-based flow control must be used between the
two application endpoints, to guarantee that the communication occurring on a
transient channel will not block the transmission of other streaming messages.

% because it is closer to the channel philosophy and HLS way of writing
% programs: if the receiver is ready to receive, a message element is sent at
% each clock-cycle and the send is actually pipelined.

For streaming collective operations, even with sufficiently large
buffers, we cannot rely on backpressure and flow control alone to coordinate
senders and receivers.  With streaming messages, we exploit that data is
produced, communicated, and consumed in a pipelined fashion, such that we can
rely on small intermediate storage in the FPGA fast memory to buffer parts of
the message during computation.  However, when data can arrive from a dynamic
number of other ranks to a single root FPGA (all-to-one), or when multiple
collectives are used in succession, some ranks can run ahead of others. In these
scenarios, data can arrive at the receiver side (the root in all-to-one, or any
rank in one-to-all) in arbitrary order.  Because of limited buffer space, the
root cannot reorder the data for a dynamic number of ranks and number of
elements. Consider, for example, a \texttt{Gather} without any coordination:
rank $i+1$ could send its full contribution to the root before rank $i$, which
the root would be unable to reorder for arbitrary message sizes.
%, and fill up the intermediate buffers in the network. 
%

\begin{figure}[tp]
	\includegraphics[width=\columnwidth]{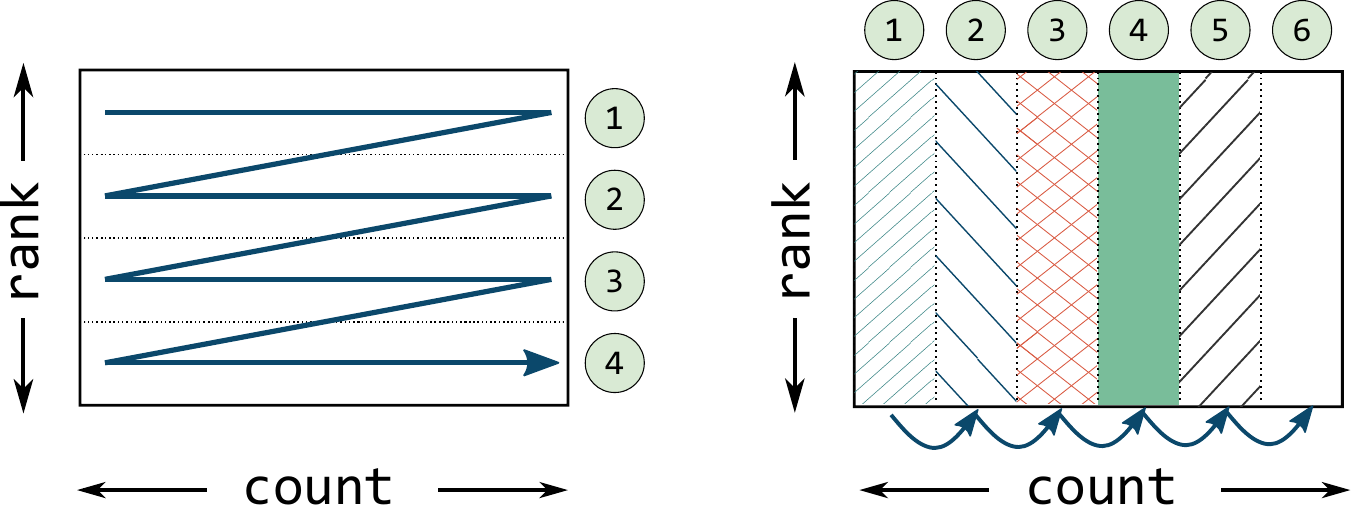}
	\vspace{-1em}
	\caption{Order of data elements communication (arrows) and coordination steps
			(numbers) for \texttt{Scatter}/\texttt{Gather} and \texttt{Reduce},
			respectively.}
	\label{fig:rendezvous}
\end{figure}

To ensure correctness in collective primitives, we employ different
synchronization protocols, depending on the type of communication used.  For
\emph{one-to-all} collectives (i.e., \texttt{Bcast} and \texttt{Scatter}),
ranks must communicate to the root when they are ready to receive before the
root starts streaming data across the network, to prevent mixing of data from
subsequently opened transient channels using the same port.  For
\emph{all-to-one} collectives (\texttt{Reduce} and \texttt{Gather}), the root
rank must communicate to each source rank when it is ready to receive the given
sequence of data. For \texttt{Bcast}, \texttt{Scatter}, and \texttt{Gather},
synchronization is done once per rank, before all data elements from the given
rank can be sent. For \texttt{Reduce}, the root synchronizes with all ranks per
tile of reduced elements. This is illustrated for
\texttt{Scatter}/\texttt{Gather} and \texttt{Reduce}, respectively, in
\figref{rendezvous}.
In \texttt{Gather}/\texttt{Scatter}, each rank will send/receive
\texttt{count} elements in sequence, only when allowed by the matching rank
(i.e., the root for \texttt{Gather} or a non-root rank for \texttt{Scatter}).
The communication between the root and the different ranks are performed in
sequential order (shown with arrow and numbers in \figref{rendezvous}). For
\texttt{Reduce}, the root must receives the first sequence of element from
\emph{all} ranks (in any order, given the associativity and commutativity
properties of the reduction operation), before receiving the next sequence from
\emph{all} ranks.  All the ranks can stream their contributions in parallel
(fill columns in \figref{rendezvous}) for the current tile being reduced
(horizontal width of columns), to the root. The root communicates to all the
other ranks when they can start sending the data for the next tile. 
	
As participating in collective operations is parallel with the number of
distinct ports, \textbf{multiple collectives can perform their rendezvous and
communication concurrently}.

\section{Reference Implementation}
\label{ref:implementation}

We present a proof-of-concept implementation of SMI, where the
transport layer and all communication primitives are implemented as HLS code,
targeting the Intel FPGA SDK for OpenCL~\cite{intel_opencl}.
% Currently, only one communicator is supported.
Network connections are implemented using I/O channels in the SDK, which are
mapped to physical interfaces implemented by the \emph{board support package}
(BSP) specifying the FPGA shell, provided by the board vendor. 
% Not sure about how many details we should put here
SMI as an interface specification is platform independent, but as the transport
layer relies on many platform-specific features, we focus on the Intel
infrastructure here.

\subsection{General Architecture}

\begin{figure}[htp]
  \includegraphics[width=7cm]{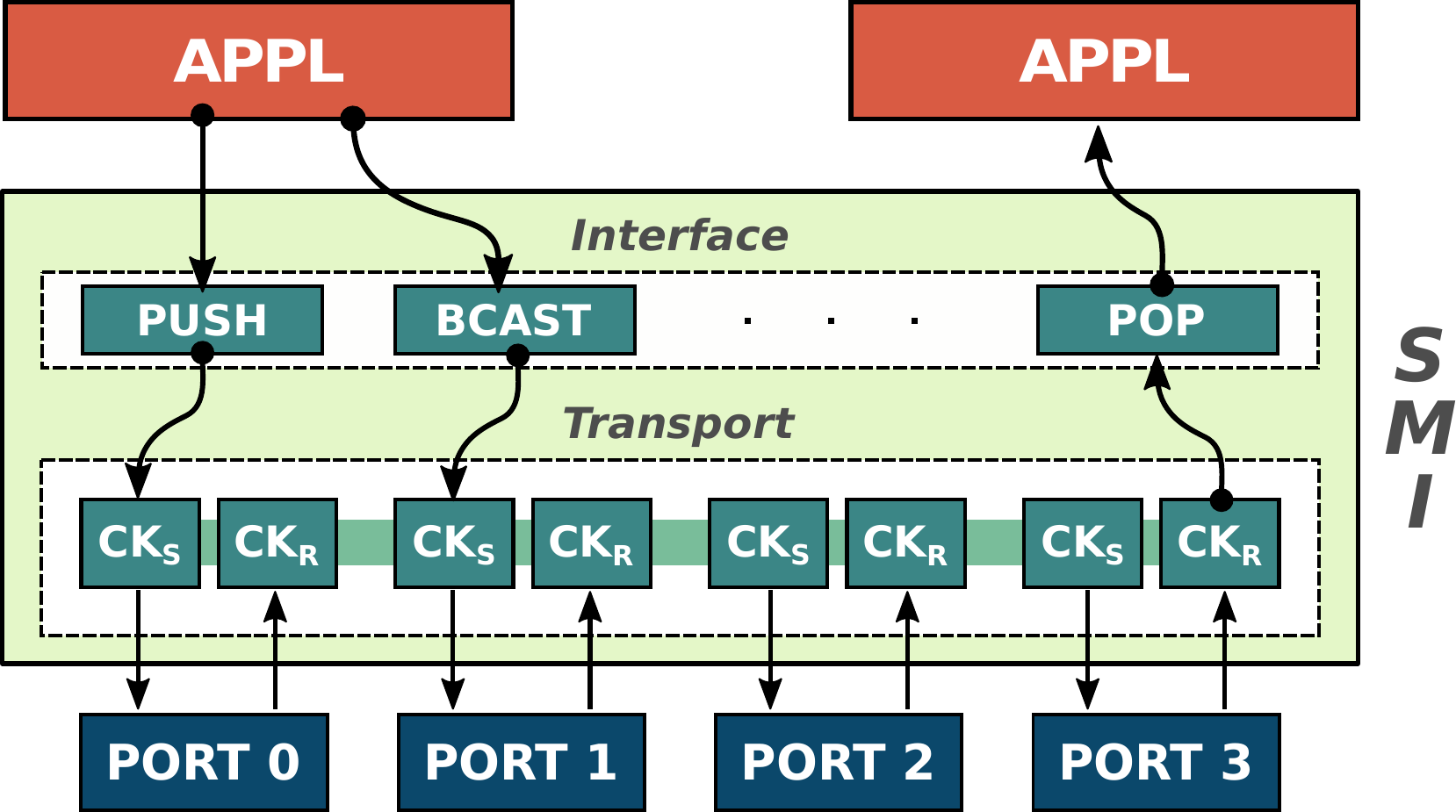}
  %\vspace{-1em}
  \caption{SMI implementation.}
  \label{fig:hls_smi}
\end{figure}

%Discuss the nnection between CKs and pplications, put a nice figure
The SMI implementation resides between applications and the network ports
exposed by the FPGA board (see \figref{hls_smi}). It is composed of two components: the
\emph{interface}, which implements the SMI primitives described in \secref{smi},
and a \textit{transport} component, which handles data transfer between endpoints.

At the SMI application interface, messages are packaged in network packets,
which have a size equal to the width of the I/O interface to the network
provided by the BSP (e.g., $\SI{32}{\Bytes}$ for the experimental platform
used). The network packet is the minimal unit of routing, and it may contain one
or more data elements.  The transport component receives network packets both
from the interface and from the network (through the BSP network interfaces).
The packet is immediately forwarded onto one of the output links according to
the specified target rank and port. The transport layer can accept one new
network packet, either coming from the network or from the applications, every
clock cycle. With the exception of the routing metadata, no bulk data needs to
be buffered in the transport layer, and the transmission of a message is fully
pipelined across the network.

\subsection{Data Forwarding}\label{sec:data_forwarding}

Each data communication to/from the network involves moving the data between
applications and the transport component through physical hardware connections
configured on the FPGA\@.  These connections are implemented using FIFO buffers,
where the internal buffer size is a compile-time parameter.
This buffer size can be tweaked according to the expected length of
the messages that will be sent, taking available FPGA resources into account. By
increasing the buffer size, a sending rank can commit more data to the network
while continuing computations, which can in some cases improve the overall
runtime. This is considered an optimization parameter, as programs must not rely
on these buffer sizes for correctness (i.e., to avoid deadlocks). 
% Johannes: uninteresting details
% By default the buffer size is set to 16, meaning that they can accommodate 16
% different network packets.}\tiziano{To be precise, the FIFO buffers accept
% Network Packets (defined later). Since each packet can hold different data
% types, the user should consider also this.}
%according to the asynchronicity degree of the channel. \johannes{Ok,
%but who sets this asynchronicity degree, and how?} \todo{}
The ports declared in \texttt{Open\_Channel} primitives are used to uniquely
identify the accessed FIFO buffer, and instructs the HLS compiler to lay down
the buffer for connecting the communication endpoint (e.g., a push or a pop) to
the transport layer. The transport component effectively acts as
\textit{middleware} between the applications and the network ports.

In the Intel FPGA SDK for OpenCL, channels are restricted to a single reader
(for input channels) or writer (for output channels): for this reason, we create
dedicated entities that handle access to the BSP network I/O channels. We refer
to these entities as \emph{send communication kernels} (\cks{}), if they send data
to the network, and \emph{receive communication kernels} (\ckr{}), if they receive
data from the network, respectively.
To perform the actual data transmission between two remote endpoints, we can
follow two approaches:
\begin{itemize}[leftmargin=*]
  \item \textit{Circuit switching}: when a \cks{} accepts the first network
  packet that composes a message, it will continue to accept data \emph{only
  from that application} until all the content of the message has been
  transferred. The message first transmits a single network packet containing
  all meta-information (source and destination rank, message data type, port,
  etc.), followed by a sequence of payload network packets.
  \item \textit{Packet switching}: \cks{} allows interleaving messages from
  different endpoints. The message is transmitted as a sequence of packets in
  which each packet must contain the meta-information necessary to route it.
\end{itemize}  
% \johannes{Since we are describing multiple options, we could consider moving
% this out of this section, which describes our specific implementation.}
The reference implementation presented here uses the second approach.	
Despite
being less bandwidth efficient, as part of each network packet is consumed by
the message header, it allow us to easily multiplex different channels, avoiding
temporary stalls due to the transmission of long messages, and all applications
can concurrently send/receive messages. 

%Ok, here is a figure for the network message in case we have empty space
%For the moment being is described in the text
%\begin{figure}
%  \includegraphics[width=6cm]{figures/network_message.pdf}
%  \caption{Network message format: OP and NE fields identify the operation type (e.g. send/receive) and the number of data items in the packet.}
%  \label{fig:network_message}
%\end{figure}

Concretely, network packets in our implementation are composed of
$\SI{4}{\Bytes}$ of header data, and a payload of $\SI{28}{\Bytes}$. The header
contains source and destination ranks ($\SI{1}{\byte}$ each), the port
($\SI{1}{\byte}$), the operation type (e.g., send/receive, $\SI{3}{\bits}$), and
the number of valid data items contained in the payload ($\SI{5}{\bits}$). We
thus truncate the rank and port information with respect to the SMI interface to
$\SI{8}{\bit}$ each to mitigate the penalty of packet switching.

Packing and unpacking network packets is implemented in the \push and \pop
primitives. \push internally accumulates data items until a network packet is
full. The packet is then forwarded to \cks{}, which will forward it towards its
destination. 
%The programmer can alter this behaviour by using the
%\texttt{SMI\_Push\_and\_flush} operation.  Similarly, when the receiver tries to
\pop internally unpacks data returned from \ckr{}, and transmits it to the
application one element at a time, according to the specified data type.

\subsection{Routing Management}\label{sec:routing}

%CKs and CKr why are they organized in that way.
In our implementation we exploit dedicated interconnection network between FPGAs
without using additional network equipment like routers or switches. Therefore,
the transport layer is in charge of implementing the routing of the data between
any pair of ranks.

\begin{figure}
  \includegraphics[width=6cm]{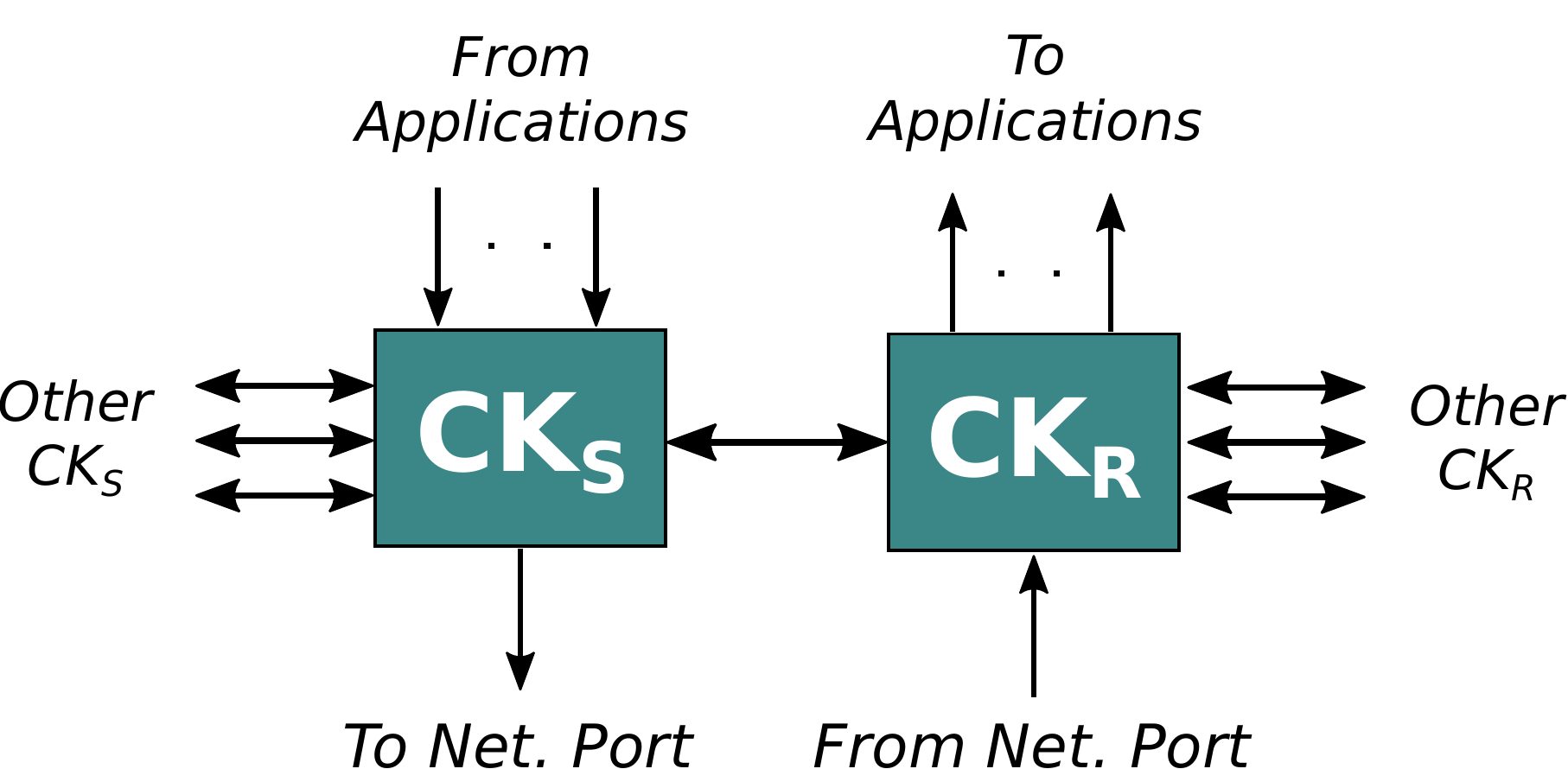}
  \vspace{-0.5em}
  \caption{Communication kernel (CK) connectivity.}
  \label{fig:ck}
\end{figure}

Each FPGA network interface is managed by a different \cks{}/\ckr{} pair. In this
way, we avoid a single centralization point that would have serialized packet
transferring.  Application endpoints are connected to one \cks{} or \ckr{} using a
FIFO buffer. The communication kernels are interconnected as shown in
\figref{ck}.
% run in an endless loop, in which they gradually try to receive a message from all the input FIFO buffers. The interconnection structure is shown in \figref{ck}. \cks{} can receive a message from its connected \ckr{}, from any other \cks{}, or from an application. \ckr{} can receive a message from its connected \cks{}, from any other \ckr{}, or from its associated networkinterface. 
After the kernel receives a packet, it consults an internal routing table to
determine where to forward the packet.  The reference implementation
employs a configurable polling scheme: when a \cks{}/\ckr{} module receives a packet
from an incoming connection, it keeps reading from the same connection up to $R$
times (where $R$ is an optimization parameter) while data is available, before
continuing to poll other ports.  With $R=1$, the \cks{} module polls a different
connection every cycle. Higher values of $R$ increase the bandwidth for
applications with a sparse communication pattern, but increases the
per-connection latency for applications where many incoming connections are
active simultaneously.

% \revision{More accurately, if a \cks{} module has $n$ connections, the maximum
% bandwidth available to \emph{a single port} is $R/n$ (when all other connections
% are inactive), and the maximum latency at \emph{a single port} is $R\cdot n$
% (when all other ports are sending data).}

The routing information used by the SMI communication kernels can be uploaded
dynamically at runtime, allowing it to be \emph{specialized} to the
interconnect, and even to the application.
We use static routing to determine the optimal paths for routing packets between
any pair of FPGAs: before the application starts, the paths between FPGAs are
computed using a deadlock-free routing scheme~\cite{deadlock-free}, according to
the target FPGA interconnection topology. 
If the interconnection topology changes, or the programs run on a different
number of FPGAs, the bitstream does not need to be rebuilt, as the routing
scheme merely needs to be recomputed and uploaded to each device.

Routing tables are buffered in on-chip memory local to each \ckr{} and \cks{}
module.
The routing tables at sender modules (\cks{}) are indexed by the destination rank
of the packet: if the destination rank is the local rank, the packet is
forwarded to the connected \ckr{}; otherwise, the packet is forwarded either to
another local \cks{} module, or to the associated network interface.
At a receiver module (\ckr{}), if the destination rank is not the local rank, it
is forwarded to the associated \cks{} module. This situation could occur when the
local rank is an intermediate hop in the route to reach  the destination.
Otherwise, the \ckr{} will use the port of the packet as an index into its routing
table. The table instructs it to either send the packet directly to a connected
application, or to forward the packet to the \ckr{} that is directly connected to
the destination port. 

By implementing the routing logic in this way, we guarantee that a rank is
reachable from all others, even if there is no physical direct connection
between them, and we allow the communication topology to be changed without
regenerating the FPGA bitstream.

\subsection{Collective Implementation}

Collective communication requires coordination between involved ranks (see
\secref{collectives}). In our reference implementation, collectives are
implemented using a simple linear scheme. The implemented SMI transport layer
uses a \textit{support kernel} for coordinating each collective.  Support
kernels reside between the application and the associated \ckr{}/\cks{} modules, and
their logic is specialized to the specific collective. \cameraready{For this reason they can also be exploited to offer different implementations of collectives, such as tree-based schema for \texttt{Bcast} and \texttt{Reduce}.} Both the root and
non-root behavior is instantiated at every rank, to allow the root rank to be
specified dynamically. For \texttt{Bcast} and \texttt{Scatter}, the support
kernel will wait at the root for the notification that a receiving rank is ready
to receive before streaming data towards it.  For \texttt{Gather}, the root rank
has to receive the data from the ranks in the correct order, which is
coordinated by the support kernel.  For \texttt{Reduce}, the support kernel will
be in charge of receiving the elements to reduce, and applying the relevant
reduction operation.  The latter implements rendezvous with a credit-based flow
control algorithm with $C$ credits, corresponding to an internal buffer of size
$C$ at the root rank holding accumulation results.  When $C$ contributions have
been received from each rank, the reduced result is forwarded to the
application, and new credits are sent to the ranks ($C$ can be considered a tile
size of the \texttt{Reduce} communication, corresponding to the width of columns
in \figref{rendezvous}).

\subsection{Development Workflow}

\begin{figure}
  \includegraphics[width=\columnwidth]{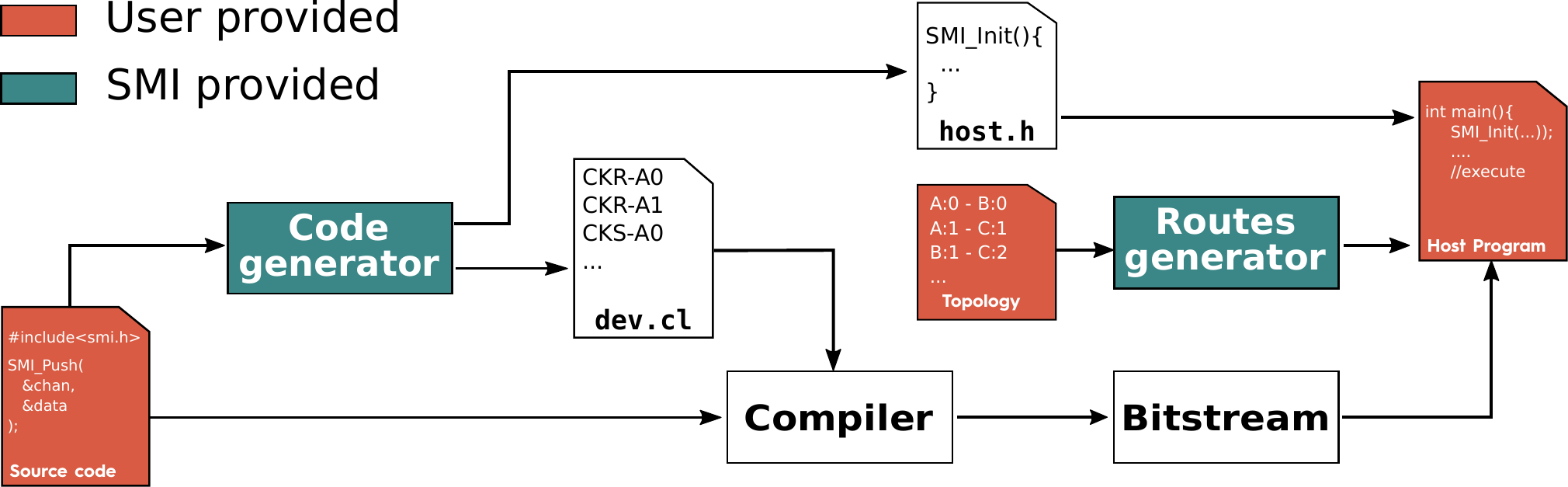}
  \vspace{-2em}
  \caption{Development workflow.}
  \label{fig:workflow}
\end{figure}
The development workflow for using SMI is depicted on \figref{workflow}. The communication logic of SMI is 
produced by a \textit{code generator}. It takes the description of SMI operations (ports, data types) as an input
and outputs a device source file with all the necessary \cks{}, \ckr{}, communication primitives and collective 
support kernel implementations that are tailored for the specified set of SMI operations. The code generator 
also outputs a host header file that contains support functions for SMI initialization.

To generate the correct input to the code generator, we provide a
\textit{metadata extractor}, that parses the user's device code with Clang,
finds all used SMI operations and extracts their metadata to a file. After the
code generator is executed on this metadata, the code-generated SMI
implementation can be compiled together with the the user's code by an FPGA
compiler to produce a bitstream. For SPMD programs, \emph{only one instance
of the code is generated}, and thus the user only needs to build a single
bitstream for any number of nodes in a multi-FPGA system.

A \textit{route generator} accepts the network topology of the FPGA cluster and
produces the necessary routing tables that drive the forwarding logic at
runtime. The topology is provided as a JSON file, which describes connections
between FPGA network ports. The route generator needs to access metadata created
by the code generator, but it doesn't modify or create any source code and
therefore it can be executed independently from the compilation (crucially, you
can change the routes without recompiling the bitstream).

Finally, the user host program takes the compiled bitstream and the routing
tables as inputs, and uses functions provided by the generated host header to
setup the routing tables, and to start all of the SMI transport components on
the FPGA.
We also provide build system integration for CMake which fully automates the
full workflow with a single function invocation.

\subsection{Implementation Portability}

The proof-of-concept implementation of SMI discussed here targets the Intel FPGA
SDK for OpenCL, but as SMI is a platform independent interface specification, it
can be implemented for other vendors, such as Xilinx FPGAs, as well. The
interface, and key concepts of the transport component can be reused, adapting
it to the target platform and SDK (changing the pragma style, FIFO buffer
management, etc.). However, the current implementation exploits Intel OpenCL I/O
channels to perform communications using the on-board network interfaces. To the
best of our knowledge, other vendors do not expose similar high level network
interfaces directly from the shell to the HLS programmer. Therefore, additional
IP cores would be necessary to port the transport component.

\section{Evaluation}
\label{sec:evaluation}

To analyze the expressiveness of SMI and the performance of our reference
implementation, we implement four microbenchmarks and two distributed
applications, showing both the SPMD and MPMD approaches of writing SMI-based
kernels.

\subsection{Experimental Setup}
\label{sec:experimental_setup}

We target the Noctua cluster at the University of Paderborn, which contains
Nallatech 520N boards, each carrying a Stratix~10 GX2800 FPGA chip. The board
exposes 4 \emph{quad small form-factor pluggable} (QSFP) transceivers as network
ports, each rated at $\SI{40}{\giga\bit\per\second}$.  The QSFP interfaces do
not implement a full reliable network stack, but implement error
correction, flow control, and handle backpressure, which we can rely on in our
communication layer.  We target the \texttt{18.1.1\_max} BSP provided by
Nallatech, which exposes the QSFP ports as 8 I/O channels (4 input and 4 output)
per device.  The I/O channels exposed to HLS are 256 bit wide, and can be
accessed using \texttt{read}/\texttt{write} primitives. \emph{All} hardware
kernels (applications and transport layer) running on the device is implemented
in OpenCL, and are compiled with the Intel Quartus~Prime~Pro 18.1.1 toolset. 

Within the target cluster, each node contains two FPGA devices, and the QSFP
ports of different FPGAs are directly connected to each other (either within or
between nodes). The FPGA interconnection topology is described by a list of
point-to-point connections, which is used to generate the routing tables. For
the experiments performed here, we had access to 8 FPGAs connected in a 2D
torus, such that all the 4 QSFP ports in each FPGA are wired to 4 distinct other
FPGAs.  Each host node is equipped with two Intel Xeon Gold 6148F CPUs, for a
total of 40 cores operating at $\SI{2.4}{\giga\hz}$, and have
$\SI{192}{\giga\byte}$ of DRAM. The nodes are interconnected using an Intel
Omni-Path $\SI{100}{\giga\bit/\s}$ network. Host code is compiled using gcc v7.3
and OpenMPI v3.1.

All experiments were executed multiple times until 99\% confidence interval is
within 5\% of the measured median. For the tests in which there is no host
intervention, few runs were sufficient to meet this condition, due to the highly
deterministic nature of FPGA codes.  Then, median times have been considered for
producing the reported performance figures.

\subsection{FPGA Resource Utilization}\label{sec:resources}
\tabref{resources} shows the FPGA resource consumption of SMI, in terms of
\textit{lookup tables} (LUTs), \textit{flip-flops} (FFs) and \textit{on-chip
memory blocks} (M20Ks).
The table reports resources consumed by the interconnection structure (\emph{Interconn.}) and communication kernels (\emph{C.K.}) both in absolute values and in fractions of the total resource capacity.
We consider two scenarios:  one where only a single network port is used, and
one where all the four available network ports are utilized. In the former case, only one pair of
communication kernels is deployed. In the latter, 4 \cks{}/\ckr{} kernels are used,
leading to additional interconnect logic. In either case, we consider one
application endpoint attached per \cks{}/\ckr{} pair.  
\begin{table}[ht]
	\small
	\centering
	\resizebox{\columnwidth}{!}{
	\begin{tabular}{>{\centering}p{1cm}rrrrrr}
		\toprule
		& \multicolumn{3}{c}{\textbf{1 QSFP }} & \multicolumn{3}{c}{\textbf{4 QSFPs}}\\
		\cmidrule(l{1pt}r{1pt}){2-4} \cmidrule(l{1pt}r{1pt}){5-7}
		 & \textbf{LUTs} & \textbf{FFs} & \textbf{M20Ks} & \textbf{LUTs} & \textbf{FFs} & \textbf{M20Ks} \\
		\midrule

		Interconn. 	& 	144 &	4,872	&	0	&	1,152	&	39,264	&	0	\\
		C. K.	&	6,186	&	7,189	&	10	&	30,960	&	31,072	&	40\\
		\midrule
		\%~of~max	&	0.3\%	&	0.7\%	&	0\%	&	1.7\%	&	1.9\%	&	0.3\%\\
	
			\bottomrule
	\end{tabular}
}
	\caption{SMI resource consumption.}
	\label{tab:resources}
	\vspace{-2em}
\end{table}

The the number of used resources grows slightly faster than linear.  This is due
to the fact that the number of input/output channels that the communication
kernels must handle increases with the number of used QSFPs. In all cases, the
resource overhead of SMI is insignificant, amounting to less than $2\%$ of the
total chip resources.
\tabref{resources_collectives} reports the resource consumption of the support
kernels used to implement the collectives evaluated in the following. These
numbers are for 32-bit floating point data, and with \texttt{SUM} as the
\texttt{Reduce} operation.
\begin{table}[ht]
	\small
	\centering
		\resizebox{\columnwidth}{!}{
	\begin{tabular}{>{\centering}p{2.5cm}rrrr}
		\toprule
		& \textbf{LUTs} & \textbf{FFs} & \textbf{M20Ks}	&\textbf{DSPs}\\
		\midrule
		Broadcast 	& 	2,560 (0.1\%) &	3,593 (0.1\%)	&	0 (0\%)	&	0 (0\%)\\
		Reduce (FP32 \texttt{SUM})		&	10,268 (0.6\%)	&	14,648 (0.4\%)	&	0 (0\%)	& 6 (0.1\%)\\
		\bottomrule
	\end{tabular}
}
	\caption{Collectives kernel resource consumption.}
	\label{tab:resources_collectives}
	\vspace{-2em}
\end{table}

%The M20K blocks are used to implement the connection to reach the DRAM, which is used by CKs to load their routing tables

\subsection{Microbenchmarks}
\label{sec:microbenchmarks}

To measure how well our reference implementation can exploit the experimental
setup, we evaluate its key characteristics by using a set of four
microbenchmarks. Communication kernels use $R=8$, and an eager
transmission protocol is used for point-to-point communication.
%\tiziano{I didn't mention the exact running frequency of all the designs. In case we can add it.}

{ 
\subsubsection{Bandwidth}
In this benchmark, a source application streams a large message to a receiver.
To test our routing approach, and measure the properties of SMI on less
connected topologies, we vary our connection topology so that the two
applications are at different network distances (hops), by disabling other
connections as needed. This is done by changing the connection list used to
compute the routes, so that the 8 FPGAs are treated as being organized along a
linear bus, rather than in a torus (without rebuilding the bitstream).

As a reference comparison for the SMI bandwidth, we consider a data transfer
performed through the host stack, where the application writes the message into
off-chip DRAM on the device, transfers it across PCIe to the host, sends it to
the remote host using an \texttt{MPI\_Send} primitive. On the receiving host,
symmetric operations are performed. \figref{bandwidth} shows the achieved
bandwidth by varying the message size and considering only the payload as data
exchanged. 
SMI approaches $91\%$ of the peak bandwidth offered by the QSFP connection,
which is $\SI{35}{\giga\bit/\s}$ when taking the $\SI{4}{\byte}$ header of each
network into account. Because the message is streamed, larger network distance
(in the absence of contention in the network) does not affect the achieved
bandwidth.  Despite using a higher bandwidth interconnect, the host-based
implementation achieves approximately one third of the SMI bandwidth, due to the
long sequence of copies through local device memory, local PCIe, host network,
remote PCIe, and remote device memory.

While this benchmark shows the bandwidth advantage on the tested
PCIe-attached setup, SMI is not coupled to a specific transport layer. For
example, in FPGAs with a high bandwidth cache-coherency bus to the host CPU
(e.g., Intel~HARP devices), or where a NIC can write to FPGA memory directly, it
could be more beneficial to use the general purpose interconnect as the
transport backend instead.

\begin{figure}[t]
	\includegraphics[width=\columnwidth]{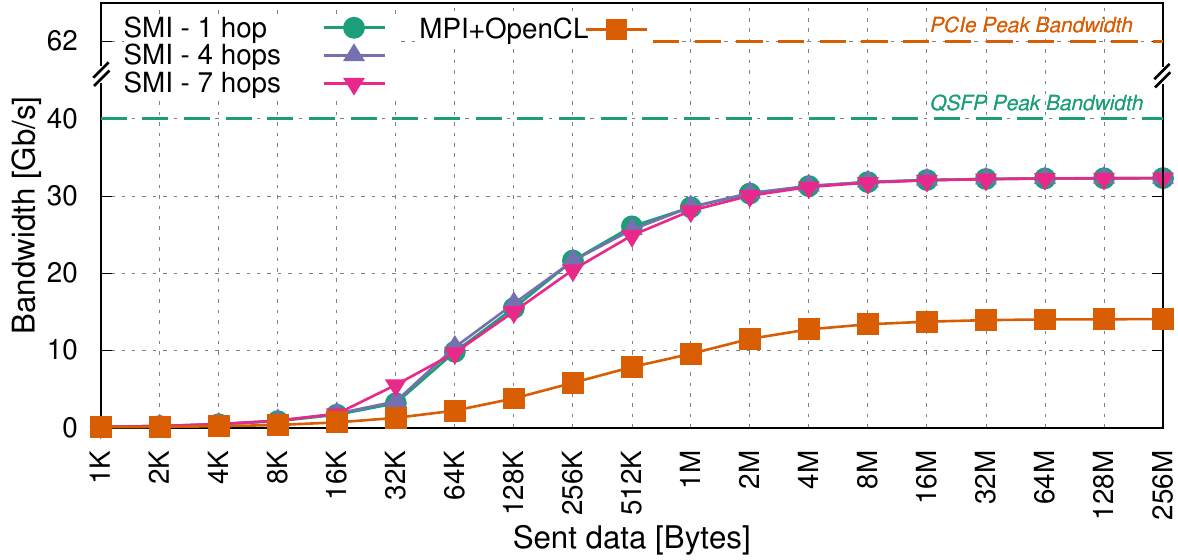}
  \vspace{-2em}
  \caption{Bandwidth comparison between SMI and host-based communication. Higher
  is better. Dashed lines indicate peak QSFP and PCIe bandwidths.}
	\label{fig:bandwidth}
\end{figure}

\subsubsection{Latency} We measure the message latency by implementing a
ping-pong benchmark of a small message between two ranks, and measure the
latency as half the execution time of a single round-trip. \tabref{latency}
shows the measured latency.  As comparison reference we implemented the same
benchmark by using host based communications. As expected, SMI is able to obtain
a much lower latency, and the time needed to reach the target increases linearly
with the increase of the network distance.

\begin{table}[hb]
	
	\small
	\centering
	%	\resizebox{\columnwidth}{!}{
	\begin{tabular}{>{\centering}cccc}
		\toprule
		\textbf{MPI+OpenCL} &  \textbf{SMI-1} & \textbf{SMI-4} & \textbf{SMI-7} \\
		\midrule
		36.61 &	0.801	&	2.896	&	5.103	\\
		\bottomrule
	\end{tabular}
	
	%}
	\caption{Measured latency in $\mu$secs. For SMI, numbers indicate the hops needed to reach destination.}
	\label{tab:latency}
	\vspace{-2em}
\end{table}

\subsubsection{Injection rate}
We measure the number of cycles that pass before a \cks{} (or \ckr{}) is able to
inject another request from the same application endpoint: i.e., the injection
\emph{latency} of the design. The injection \emph{rate} is computed from this
according to the clock frequency of the design. To benchmark this, we use a
sender application that opens a send channel and sends a message with one element at
\emph{every iteration} of a pipelined loop (i.e., every clock cycle).
Independent of the network, the sender is thus capable of an injection rate
equivalent to the clock frequency of the design.

We benchmark the scenario in which we have 4 communication channels per FPGA
with 4 \ckr{}/\cks{} pairs and we let the parameter $R$ vary. We measure the injection rate by dividing the number of
injected messages by the kernel execution time, then multiplying this by the clock
frequency to obtain the injection latency.
\begin{table}[htb]
	\small
	\centering
	%	\resizebox{\columnwidth}{!}{
	\begin{tabular}{>{\centering}cccc}
		\toprule
		$\mathbf{R=1}$ &  	$\mathbf{R=4}$ & 	$\mathbf{R=8}$ & 	$\mathbf{R=16}$\\
		\midrule
		5 &	2.5	&	1.8	&	1.69	\\
		\bottomrule
	\end{tabular}
	%}
  \caption{Average injection rate in number of cycles.}
	\label{tab:injection}
	\vspace{-2em}
\end{table}

For the case in which $R=1$ we 
measure $5$ clock cycles. This latency is due to the implemented packet
switching protocol (see \secref{routing}), where the \cks{} module polls
a different port at every cycle, corresponding to a latency of 5 cycles (1 from the application, 1 from \ckr{}, 3 from
other \cks{} modules). As long as $R$ increases, the injection rate decreases as the communication kernels will spend more time in reading from the same port.
% Note that, in any case, the communications kernels are able to handle a
% network packet coming from other sources, like other applications or
% communication kernels.
%In general, given the number of QSFP/network ports $Q$, and the number of
%application endpoints $A$, the injection latency when considering the full
%network will be $Q+\ceil{A/Q}$ clock cycles.

\subsubsection{Collective operations}
We benchmark the time required to broadcast and reduce a message of
varying size between 4 and 8 FPGAs, considering two different connection
topologies: a torus, and a linear bus. The evaluation is done with 32-bit floating point
data and with \texttt{SUM} as the \texttt{Reduce} operation.  Results are shown
in \figref{broadcast} and \figref{reduce} for broadcast and
reduce, respectively.  For broadcast, SMI is able to achieve
lower communication time for all the considered input sizes. SMI achieves similar performance independently of the considered connection topology. For small to medium-sized messages, SMI's \texttt{Reduce} outperforms going over the host using OpenCL and MPI, but loses its benefit at high message sizes.
The credit-based flow control algorithm implemented by the root is latency sensitive, therefore the time to completion increases with the diameter of the network. The SMI
reference implementation does not yet implement tree-based collectives,
resulting in a higher congestion in the root rank.

\begin{figure}[htb]
	\includegraphics[width=\columnwidth]{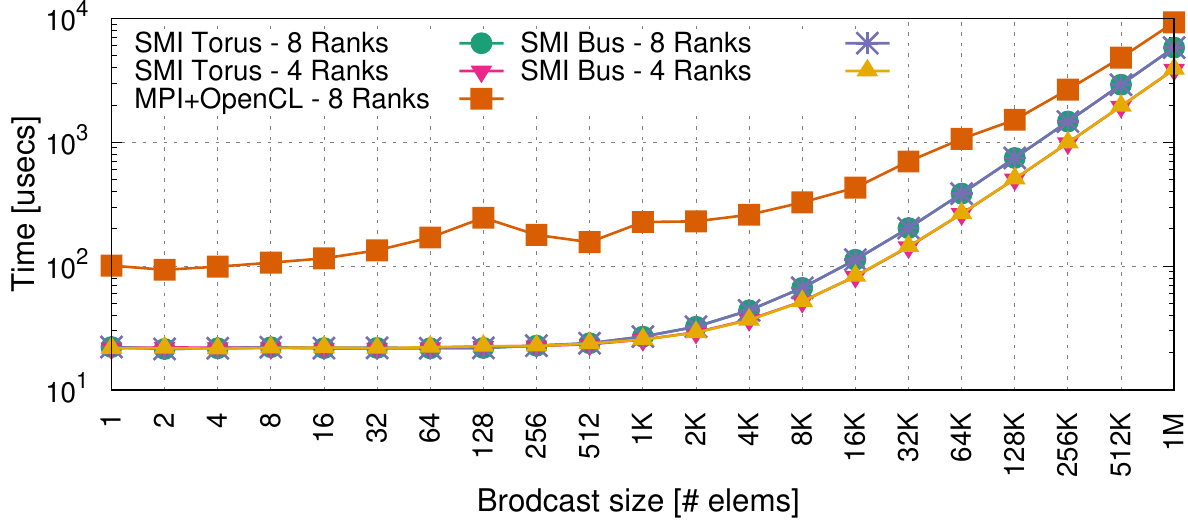}
  \vspace{-2em}
  \caption{\texttt{Bcast} benchmark comparison between SMI and
  host-based communication. Lower is better.}
	\label{fig:broadcast}
\end{figure}

\begin{figure}[htb]
	\includegraphics[width=\columnwidth]{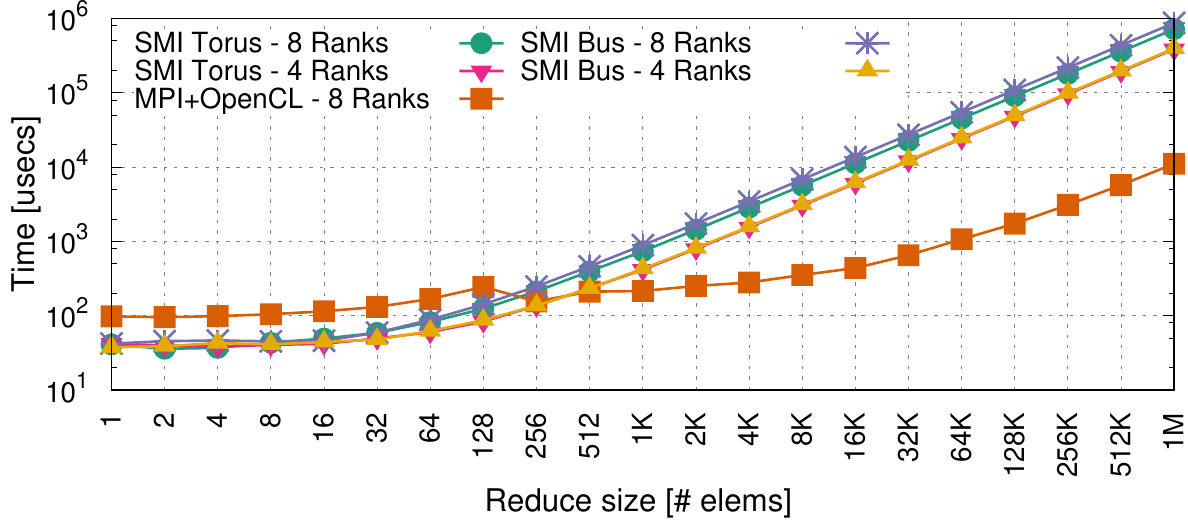}
  \vspace{-2em}
  \caption{\texttt{Reduce} benchmark comparison between SMI and
  host-based communication. Lower is better.}
	\label{fig:reduce}
\end{figure}
%For small to medium-sized messages, SMI
%outperforms going over the host using OpenCL and MPI, but loses its benefit at high message sizes, as the current SMI implementation does not yet implement tree-based collectives, resulting in a higher total communication volume.
% The further study and implementation of efficient collective communication in
% multi-FPGAs system is left as future work.
%We provided a basic implementation of the broadcast

%not able to put MPI+OpenCL vertically centered

%\textit{Results:} For the bandwidth, we achieve full bandwidth on the single connection (40Gb/s). If we want to consider only the payload bandwidth, then we have to remove $1/8th$ from that quantity (so 35Gb/s), since in our packets the header consumes 4 bytes out of 32. There is the story that I mentioned about the fact that we should avoid having a lot of \cks{}/\ckr{} if we don't use them since this will penalize us. I don't think that we should discuss this in the paper.

\subsection{Applications}
\label{sec:applications}

\subsubsection{GESUMMV}
\label{sec:gesummv}

Dense linear algebra makes up some of the most common routines in HPC
applications, and are good candidates for exploiting the spatial parallelism
offered by FPGAs. 
% Dense linear algebra computation are a good candidate to being implemented on FPGA, thanks to their exploitable spatial parallelism.
We consider the \texttt{GESUMMV} routine, which is a part of the Extended BLAS
set~\cite{extended_blas}, which in turn invokes BLAS subroutines. It computes $y
= \alpha Ax + \beta Bx$, where $\alpha$ and $\beta$ are scalars, $x$ and $y$ are
vectors of length $N$, and $A$ and $B$ are matrices of size $N\times N$. 
To show the benefit of SMI, we implement a single chip and a distributed version of the routine. 
The single-FPGA implementation consists of two matrix-vector multiplications
(\texttt{GEMV} routines) that compute in parallel, and stream their results to a
vector addition module (\texttt{AXPY} routine) producing the final result (\figref{gesummv_impl}, left). As these
routines are memory-bound, the computation is bottlenecked by memory bandwidth.

\begin{figure}[htb]
	\includegraphics[width=\columnwidth]{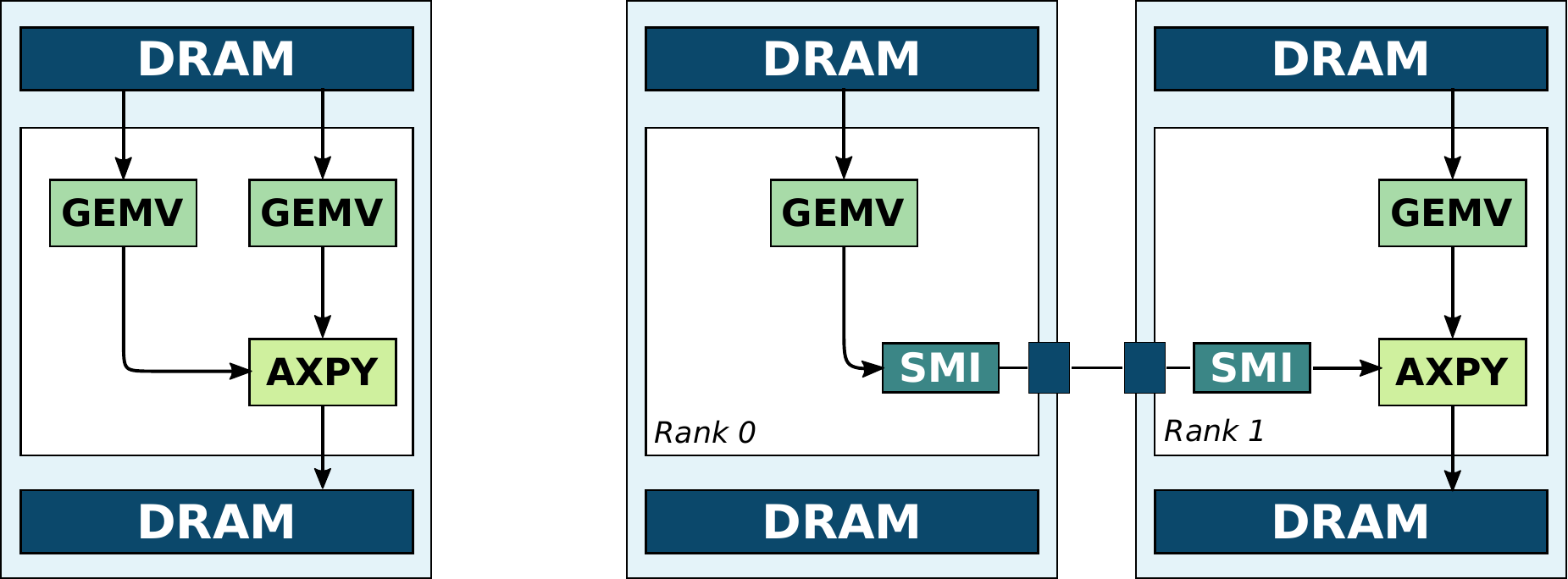}
	\vspace{-2em}
	\caption{\texttt{GESUMMV} implementations.}
	\label{fig:gesummv_impl}
\end{figure}

\noindent The distributed implementation is obtained by functional
decomposition, and it is implemented as a MPMD program using two ranks
(\figref{gesummv_impl}, right). Rank~0 computes the first matrix-vector
multiplication and sends the result elements to rank 1 using an SMI channel. On
rank~1, the second matrix-vector multiplication and the vector addition are
performed, receiving data from both local DRAM and the remote \texttt{GEMV}
routine. The full application thus gains access to twice the memory bandwidth
across the two FPGAs. The implementations of \texttt{GEMV} and \texttt{AXPY} are
derived from an open-source synthesizable library \cite{fblas}.

\figref{gesummv} shows the expected speedup of ${\sim}2\times$ of the
distributed implementation over the single-chip implementation. Execution times
of the SMI benchmarks are reported on top of the histogram boxes. Adapting the
application required only minimal code modifications to the kernel, with a
difference of 8 lines of code: \texttt{GEMV} on rank~0 is changed to perform an
SMI send rather than pushing its result to a regular FIFO, and the vector
addition is modified to read one of its inputs from an SMI network channel.

\begin{figure}[htb]
	\includegraphics[width=\columnwidth]{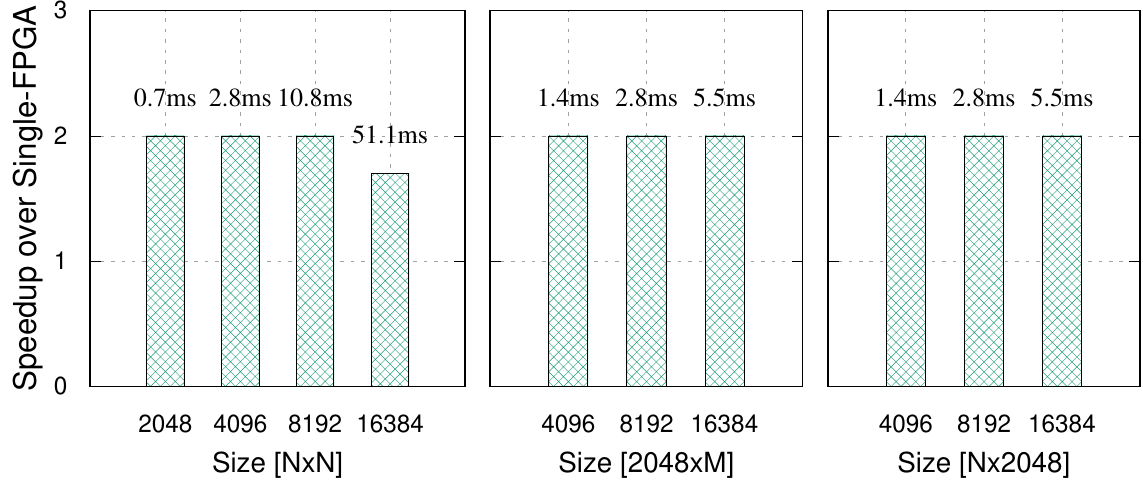}
    \vspace{-2em}
	\caption{\texttt{GESUMMV} benchmark results for different matrix sizes (square and rectangular).}
	\label{fig:gesummv}
\end{figure}

\subsubsection{Stencil}
\label{sec:stencil}

\begin{figure}[b]
  \centering
  %\vspace{1em}
  \includegraphics[width=.7\columnwidth]{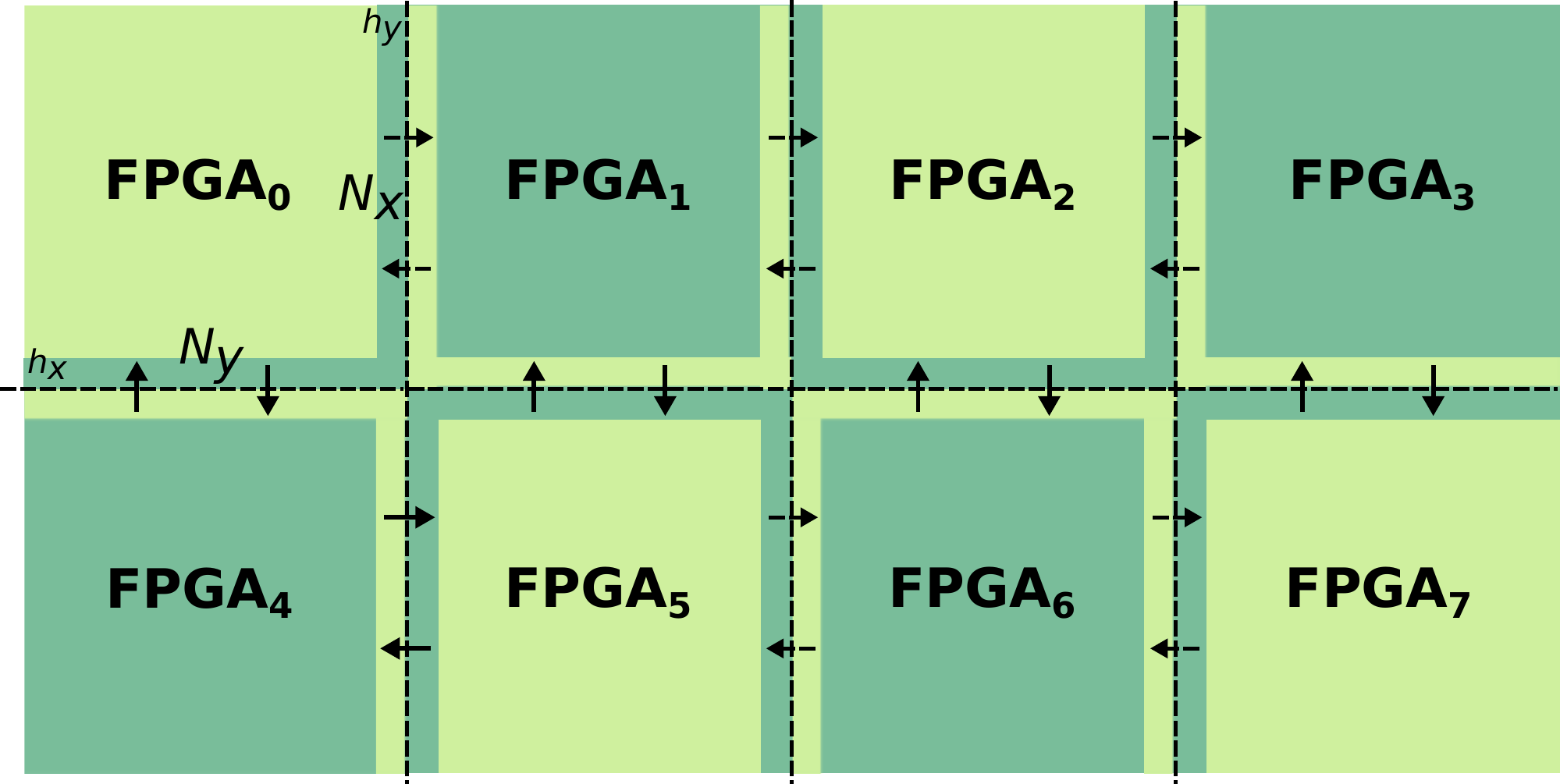}
  \vspace{-1em}
  \caption{Halo region exchange between FPGAs.}
  \label{fig:halo_exchange}
\end{figure}

Stencil applications are a suitable target for FPGA acceleration, as their
regular access pattern allows implementing memory reuse schemes that make highly
efficient use of on-chip memory. Even with perfect reuse across the spatial
domain, however, stencils generally exhibit low computational intensity.
Additional reuse can be obtained by using time tiling, which is implemented on
FPGAs by connecting a linear array of processing elements in a deep pipeline,
executing multiple timesteps in parallel~\cite{sano2014multi,
zohouri2018combined}. When parallelizing to multiple FPGAs,
Sano~et~al.~\cite{sano2014multi} simply extend this array to multiple FPGAs by
using serial connections between them in a streaming model (as conceptually
illustrated in \figref{streaming}).

For large stencil domains, FPGA designs must tile the spatial domain in addition
to the time domain, as the required buffer size grows with the size of the
domain. This results in a halo region of redundant computations, which is
proportional to the number of pipelined timesteps executed in
parallel~\cite{zohouri2018combined}. This puts a hard limit on the scalability
of this approach, as the number of redundant computations will in the extreme
case dominate ``useful'' computations. Furthermore, not all stencils require or
allow time tiling, leaving spatial parallelism (e.g., vectorization) as the only
option to speed up the computation, in which case the problem becomes memory
bound. It is thus desirable to parallelize spatially across multiple FPGAs,
exploiting both compute resources and memory bandwidth of multiple devices.

We implement a SPMD distributed memory FPGA stencil code using SMI. The spatial
domain is scattered to multiple devices before execution, and the devices
exchange halo regions during computation. Shift registers are used to achieve
perfect spatial reuse within each FPGA. We decompose the domain in two
dimensions, such that each FPGA communicates to and from a north, west, east,
and south neighbor, shown in \figref{halo_exchange}. Additional tiling
can be employed for large domains and 3D stencils by further decomposing the
domain on each rank without affecting the communication pattern. The
communication is naturally expressed with streaming messages in the pipelined
code. A snippet of the communication code is shown in \lstref{halo_exchange}: at
each timestep, channels are opened to adjacent ranks using a distinct port for
each neighbor, and data is read from the network during computation using
\texttt{SMI\_Pop} commands. Although the west and east halos are not contiguous
in memory, they are expressed as a single message in the streaming messages
model. Due to the transient nature of SMI channels, all ranks will be configured
with the same bitstream, and the rank of adjacent neighbors is computed at
runtime. If no neighbor exists (e.g., the west and north neighbor for
$\text{FPGA}_0$ in \figref{halo_exchange}), the given channel simply remains
unused. 

\begin{figure}[b]
  \begin{minted}{C++}
for (int t = 0; t < T; t++) {
  int num_elems = h_y*(N_x-2*h_x); // Size of halo region 
  int r_x = rank / RY;  // Rank coordinates
  int r_y = rank % RY;
  SMI_Channel recv_west = SMI_Open_recv_channel(
    num_elems, SMI_FLOAT, r_x * RY + (r_y - 1), 1, 
    SMI_COMM_WORLD);
  SMI_Channel recv_east = SMI_Open_recv_channel(
    num_elems, SMI_FLOAT, r_x * RY + (r_y + 1), 2, 
    SMI_COMM_WORLD);
  // ...open remaining channels...
  for (int i = 0; i < N_x; i++) {    // Pipelined
    for (int j = 0; j < N_y; j++) {  // region
      float value;
      bool on_corner = /* ... */;
      if (r_y > 0 && j < h_y && !on_corner) {  // On left  
        SMI_Pop(&recv_west, &value);           // halo
      } else if (ry < RY - 1 && j >= N_y - h_y && 
            !on_corner) { 
            SMI_Pop(&recv_east, &value);
        // ...handle other halos and boundary conditions...
      } else
        value = memory[i*N_y + j];
      write_channel_intel(to_kernel, value); // Stream to 
} } }                                        // compute
  \end{minted}
  \vspace{-1em}
  \captionof{listing}{Communication in pipelined stencil code.}
  \label{lst:halo_exchange}
\end{figure}

To fully hide communication, the communication volume of the non-halo region of
size $(N_x - 2 h_x) \cdot (N_y - 2 h_y)$ must be greater than the communication
volume of the halo regions of size $2 \cdot (2 h_x N_y + 2 h_y N_x)$ (send and
receive), weighted by the memory bandwidth consumed to read values from memory
on each FPGA ($B_\text{mem}$) and the network bandwidth between two adjacent
FPGAs ($B_\text{comm}$), respectively (for our system, we consider
$B_\text{comm}$ constant. In larger networks, $B_\text{comm}$ depends on how
ranks are mapped to the network topology). That is, the following inequality
must hold:
\begin{align*}
  \frac{(N_x - 2 h_x) \cdot (N_y - 2 h_y)}{B_\text{mem}} \geq
  \frac{4\left(N_x \cdot h_y + N_y \cdot h_x\right)}{B_\text{comm}}
\end{align*}
Since the left-hand side grows quadratically with the stencil domain size, this
condition is easily met when tackling large problems. 

For benchmarks we use a 4-point stencil (i.e., $h_x = h_y = 1$). 
We demonstrate the benefit of spatial tiling in a distributed memory FPGA
setting using SMI by showing the strong scaling behavior of five kernels
executed over the same stencil domain: a vectorized kernel with perfect spatial
reuse, reading $16$ elements per cycle from a single DDR bank (1 bank/1 FPGA); a
spatially tiled kernel running on a single node, reading $64$ elements per cycle
across all four memory modules of the FPGA (4 banks/1 FPGA); an SMI code running
on four FPGAs, each reading $16$ elements per cycle from a single memory bank
per FPGA (1 bank/4 FPGAs); an SMI code running on four FPGAs, each reading $64$
elements per cycle across all memory banks (4 banks/4 FPGAs); and an
SMI implementation running on 8 FPGAs organized in a $2\times 4$ shape, each
reading $64$ elements per cycle across all memory banks (4 banks/8 FPGAs).
Results are shown in \figref{stencil_benchmark} for a $4096\times4096$ domain,
executed for 32 timesteps using the torus connection topology. We
executed the same benchmarks with the FPGAs organized in a bus topology, and
observed this to not affect the execution time.

Exploiting four banks on a single FPGA, and exploiting one bank per FPGA on four
FPGAs, both show a nearly identical speedup of $3.5\times$, demonstrating that
communication and computation is fully overlapped. When reading $64$ elements on
four FPGAs, we get the exact product $3.5 \cdot 3.5 = 12.3\times$ as speedup
over the single bank version, while 8 FPGAs exhibit a speed of
$23.1$. In \figref{stencil_benchmark2} we evaluate weak scaling, by
reporting the average computation time per grid point obtained with different
grid sizes for the 4 and 8 FPGAs setups. At large grid sizes, 8 FPGAs achieve a
2x speedup over 4 FPGAs.}

\begin{figure}
  \includegraphics[width=\columnwidth]{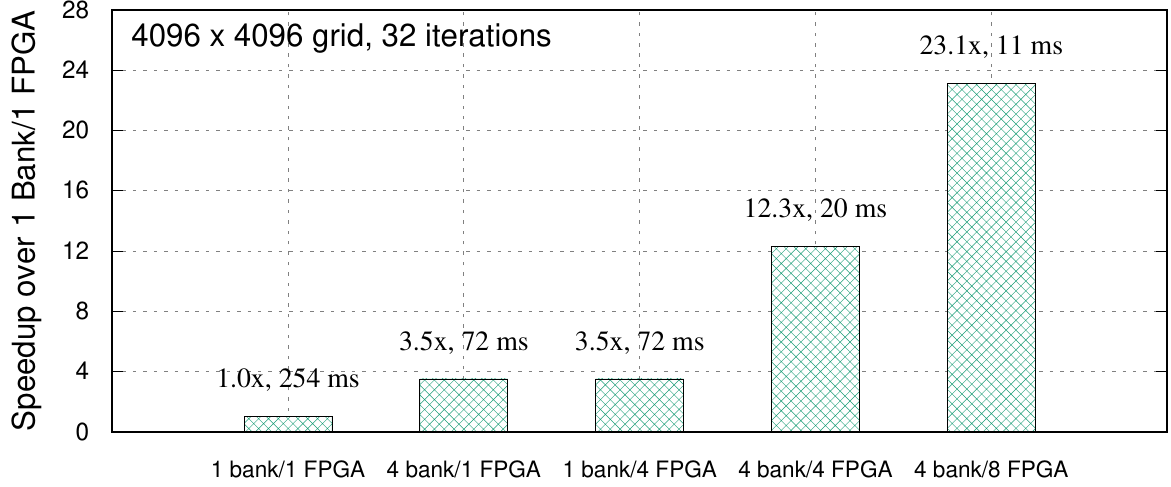}
  \vspace{-2em}
  \caption{Stencil benchmark with and without SMI.}
  \label{fig:stencil_benchmark}
\end{figure}
\begin{figure}
	\includegraphics[width=\columnwidth]{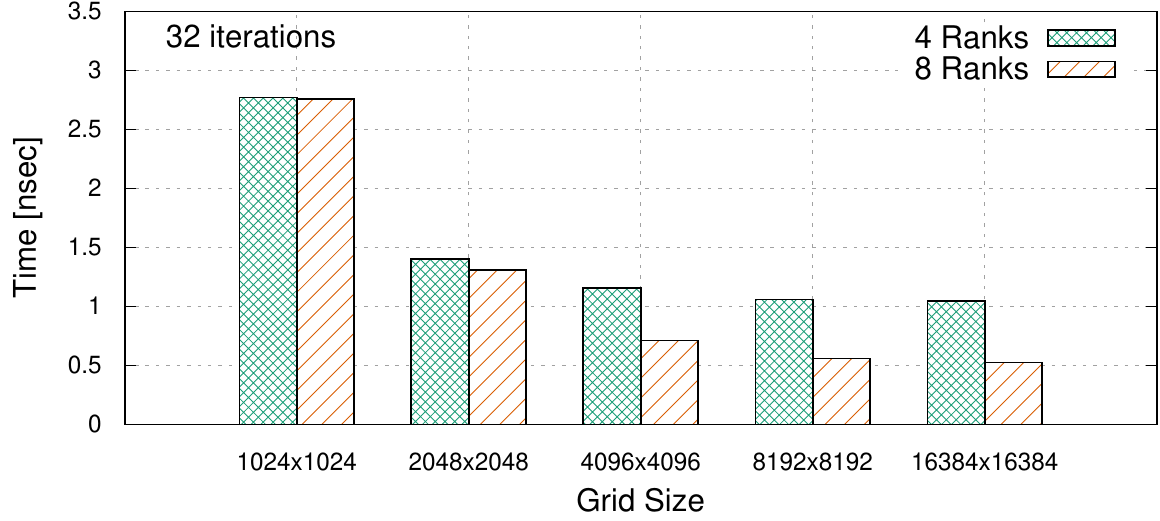}
	\vspace{-2.5em}
  \caption{Average execution time per stencil point of SMI for varying
  grid size with 4 memory banks per FPGA.}
	\label{fig:stencil_benchmark2}
\end{figure}

\vspace{1em}
With SMI and our reference implementation, we show that we can execute FPGA
programs \emph{in both MPMD and SPMD fashion}, target \emph{any network
topology}, specialize to the target network topology, and \emph{scale the number
of FPGAs using the \textbf{same} bitstream}.  Adapting to SMI requires minimal
code intervention, as the interface integrates into the conventional streaming
approach taken by pipelined HLS codes, and is thus nearly equivalent to
parallelizing the code on a single device.

% RIP
% \subsubsection{Another one}
% \textit{Comments:}
% Possible candidates are:
% \begin{itemize}
%   \item one application of the Rodinia benchmark, for example NW which is a dynamic programming application.
% \end{itemize}

%
%
%\revision{
%	\section{Discussion}
%	
%Implementing streaming messages opens different challenges that have to be addressed. Some of them are related to the communication models (streaming vs bulk transfers), others are due to the particular hardware that we are using.
%	
%In this section we should argue about some of our choices and explain why these are not dictated by implementation decisions.
%\begin{itemize}
%	\item rendezvous for point-to-point: in theory the current implementation could deadlock. We can say that we use deadlock-free routing schema and still using an eager protocol for P2P communications;
%	\item rendezvous for collective: we need differend kind of rendezvous protocol for different collectives. For the reduce we need to coordinate for each reduced elements. For gather we just need an ordering between the ranks. This rendezvous is needed because we are assuming that we have limited buffer space;
%	\item we have a constant channel width, so we can't unroll communications as we want.
%	\item efficient collective implementation
%\end{itemize}
%}
\section{Related Work}
\label{sec:related_work}

%To check also this \textit{VerilogCSP} (High Level Modeling of Channel-Based
%Asynchronous Circuits Using Verilog )
In previous work targeting multi-FPGA systems, FPGAs interconnected by
point-to-point serial connections are typically programmed according to the
streaming model (\figref{streaming}).  The common approach to scaling to
multiple FPGAs is to organize the computation in a pipeline spanning across
multiple chips, in which each stage communicates only with the previous and the
successive stage (e.g., systolic array approaches).
% Distributed application  use cases
%Sato
Sato~et.~al.~\cite{sano2014multi} parallelize 2D and 3D stencils in this way by
pipelining a linear array of processing elements across multiple FPGAs. Each
processing element performs one time-step, and the results are streamed to the
next one.
Zhang~et~al.~\cite{cnn_cluster_fpga} propose an implementation of a
convolutional neural network that pipelines 6 FPGA devices connected in a ring
topology. In their solution, a network layer is implemented by a single stage.
Geng~et~al.~\cite{fdeep} addresses a similar problem by proposing a
pipelined implementation in which layers are distributed across multiple
pipeline stages.
Owaida~et~al.~\cite{inference_tree} parallelize the inference over a decision
tree ensemble. They proposed a communication shell to implement communications
between FPGAs either by using serial links or via host intervention.
In all the aforementioned cases, the application programmer is limited to the
streaming model in expressing a distributed computation, constructing the exact
path to move the data across the FPGAs in the system.
With streaming messages and SMI we propose a more flexible solution, allowing
programmers to dynamically exploit arbitrary communication topologies in the
spirit of MPI, while maintaining a streaming programming model for computations.

%% MPI on accelerators
Despite accelerators being ubiquitous in supercomputers and data centers, there
is no unified programming model or library for communicating directly from/to
accelerator devices.  Traditionally, programmers have been forced to use a mix
of different programming models (e.g, MPI+CUDA or MPI+OpenCL).  More recently,
there has been an effort from the HPC community in developing programming models
and libraries that treat accelerators as first-class citizens:
%MPI-ACC
Aji~et~al.~\cite{mpi-acc} propose MPI-ACC, an accelerator-aware MPI
implementation, to support data transfers in heterogeneous clusters;
%dcuda dcuda
Gysi~et~al.~\cite{dcuda} propose dCuda, which combines the CUDA programming
model with a subset of MPI remote memory access operations;
%By over-subscribing the device, the latency of device-side implemented
%communications is masked by computation.
%impacc impacc
and the authors of IMPACC~\cite{impacc} propose integrating MPI and OpenACC, by
mapping all the available host and device memories in a node to a single unified
node virtual address space.
% A message handler thread in execution on the host is used to handle the
% communications from/to the device.
All these solutions involve intervention of the host to perform the actual data
communication. 
%This involve a sequence of Device-to-host, host-to-host, host-to-device and  operations and the execution of a proper runtime on the host node.
In SMI, we provide an accelerator-oriented communication library, and show how
this can be used to exploit a dedicated FPGA interconnect, avoiding costly trips
through the host nodes, saving PCIe, host DRAM, and host network bandwidth.
% Other candidates
% - starpu-mpi ?
% - MPI-ACC: Accelerator-Aware MPI for Scientific Applications

%% Distributed FPGA library

In the context of FPGA programming, various works address applying the message
passing model to multi-FPGA systems.
%TMD-MPI
TMD-MPI~\cite{tmd-mpi} implements a subset of MPI primitives for multi-FPGA
systems. The authors implement a VHDL-based engine that performs communications,
exploiting shared memory (if the FPGAs are attached to the same host) or a
specialized network interface (for remote communications).
%there is a nice sentence in the paper in which the author wonder wheter MPI is a good model for FPGAs or no
%DUA
Shu~et~al.~\cite{dua} propose DUA, a communication architecture that provides
uniform access for FPGAs to data center resources like CPUs, GPUs, and disks.
The system is implemented in Verilog, but provides an OpenCL interface, and
targets FPGAs implementing a full network stack in a cloud setting. The system
provides basic message-passing primitives, but does not go further to address
the programming model.
%Galapagos, galapagos
Eskandari~et~al.~\cite{galapagos} propose HUMboldt, a message passing
communication layer, supporting messages to be sent among different FPGA kernels
and CPU kernels.
In all these works, authors apply the message passing model \emph{directly} to
program a multi-FPGA system. With SMI, we explicitly address the issue of
programming communication in a pipelined HLS setting, providing a model and
interface that is familiar to HPC users, yet integrates well into hardware
designs.

Finally, %Novo-G#
George et al.~\cite{novo-g-sharp} present a network infrastructure for allowing
communication among FPGAs organized in a 3D torus. In their subsequent
work~\cite{novo-g-sharp-opencl}, they build an OpenCL abstraction on top of this
network stack to enable inter-FPGA communications in HLS programs. In contrast
to SMI, their solution exploits the streaming model, without defining a precise
communication interface and with no support for collective communications.

%% Why we are cool? We merge FPGA and MPI worlds by introducing streaming messages

\section{Conclusion}
\label{sec:conclusion}

We propose \emph{streaming messages}, a communication model for distributed
memory programming on reconfigurable hardware. Streaming messages unify message
passing and traditional streaming communication, allowing transient channels to
be dynamically established between multiple FPGAs in distributed systems, while
maintaining a programming model that integrates seamlessly into HLS designs.
%The FPGA programmer can use its classical 
To capture and expose the semantics of streaming messages, we introduce SMI, a
communication interface specification for HLS programs, drawing inspiration from
MPI, but designed to fit the hardware programming model, and release an open
source reference implementation for use with OpenCL-capable Intel FPGAs. 
With the simple and powerful model offered by SMI, we hope to further the
viability of FPGAs as a HPC accelerators, and make distributed programming on
FPGAs more accessible to both HPC and hardware developers. 

\begin{acks}

We wish to thank the Paderborn Center for Parallel Computing
(PC\textsuperscript{2}), in particular Christian Plessl and Tobias Kenter, for
access, support, maintenance, and upgrades, sometimes on very short notice. We
also would like to thank Mohamed Issa (Intel Corporation), for valuable
suggestions. This project has been supported from the European Research Council
(ERC) under the European Union's Horizon 2020 programme, Grant Agreement No.
678880 (DAPP), and Grant Agreement No. 801039 (EPiGRAM-HS).  Jakub Ber\'anek was
supported by the European Science Foundation through the ``Science without
borders'' project, reg. nr.  CZ.02.2.69/0.0./0.0./16\_027/ 0008463 within the
Operational Programme Research, Development and Education.
\end{acks}

\bibliographystyle{ACM-Reference-Format}
\bibliography{FPGA_MPI}


\begin{thebibliography}{28}

%%% the Web address.

\ifx \showCODEN    \undefined \def \showCODEN     #1{\unskip}     \fi
\ifx \showDOI      \undefined \def \showDOI       #1{#1}\fi
\ifx \showISBNx    \undefined \def \showISBNx     #1{\unskip}     \fi
\ifx \showISBNxiii \undefined \def \showISBNxiii  #1{\unskip}     \fi
\ifx \showISSN     \undefined \def \showISSN      #1{\unskip}     \fi
\ifx \showLCCN     \undefined \def \showLCCN      #1{\unskip}     \fi
\ifx \shownote     \undefined \def \shownote      #1{#1}          \fi
\ifx \showarticletitle \undefined \def \showarticletitle #1{#1}   \fi
\ifx \showURL      \undefined \def \showURL       {\relax}        \fi
% invisible to TeX
\providecommand\bibfield[2]{#2}
\providecommand\bibinfo[2]{#2}
\providecommand\natexlab[1]{#1}
\providecommand\showeprint[2][]{arXiv:#2}

\bibitem[\protect\citeauthoryear{{Aji}, {Panwar}, {Ji}, {Murthy}, {Chabbi},
  {Balaji}, {Bisset}, {Dinan}, {Feng}, {Mellor-Crummey}, {Ma}, and
  {Thakur}}{{Aji} et~al\mbox{.}}{2016}]%
        {mpi-acc}
\bibfield{author}{\bibinfo{person}{A.~M. {Aji}}, \bibinfo{person}{L.~S.
  {Panwar}}, \bibinfo{person}{F. {Ji}}, \bibinfo{person}{K. {Murthy}},
  \bibinfo{person}{M. {Chabbi}}, \bibinfo{person}{P. {Balaji}},
  \bibinfo{person}{K.~R. {Bisset}}, \bibinfo{person}{J. {Dinan}},
  \bibinfo{person}{W. {Feng}}, \bibinfo{person}{J. {Mellor-Crummey}},
  \bibinfo{person}{X. {Ma}}, {and} \bibinfo{person}{R. {Thakur}}.}
  \bibinfo{year}{2016}\natexlab{}.
\newblock \showarticletitle{MPI-ACC: Accelerator-Aware MPI for Scientific
  Applications}.
\newblock \bibinfo{journal}{\emph{IEEE Transactions on Parallel and Distributed
  Systems}} \bibinfo{volume}{27}, \bibinfo{number}{5} (\bibinfo{date}{May}
  \bibinfo{year}{2016}), \bibinfo{pages}{1401--1414}.
\newblock
\showISSN{1045-9219}
\urldef\tempurl%
\url{https://doi.org/10.1109/TPDS.2015.2446479}
\showDOI{\tempurl}


\bibitem[\protect\citeauthoryear{Blackford, Demmel, Dongarra, Duff, Hammarling,
  Henry, Heroux, Kaufman, Lumsdaine, Petitet, Pozo, Remington, and
  Whaley}{Blackford et~al\mbox{.}}{2002}]%
        {extended_blas}
\bibfield{author}{\bibinfo{person}{Susan Blackford}, \bibinfo{person}{James
  Demmel}, \bibinfo{person}{Jack Dongarra}, \bibinfo{person}{Iain Duff},
  \bibinfo{person}{Sven Hammarling}, \bibinfo{person}{Greg Henry},
  \bibinfo{person}{Michael Heroux}, \bibinfo{person}{Linda Kaufman},
  \bibinfo{person}{Andrew Lumsdaine}, \bibinfo{person}{Antoine Petitet},
  \bibinfo{person}{Roldan Pozo}, \bibinfo{person}{Karin Remington}, {and}
  \bibinfo{person}{Clint Whaley}.} \bibinfo{year}{2002}\natexlab{}.
\newblock \showarticletitle{An Updated Set of Basic Linear Algebra Subprograms
  ({BLAS})}.
\newblock \bibinfo{journal}{\emph{ACM Trans. Math. Softw.}}
  \bibinfo{volume}{28}, \bibinfo{number}{2} (\bibinfo{date}{June}
  \bibinfo{year}{2002}), \bibinfo{pages}{135--151}.
\newblock
\showISSN{0098-3500}


\bibitem[\protect\citeauthoryear{{Chung}, {Fowers}, {Ovtcharov}, {Papamichael},
  {Caulfield}, {Massengill}, {Liu}, {Lo}, {Alkalay}, {Haselman}, {Abeydeera},
  {Adams}, {Angepat}, {Boehn}, {Chiou}, {Firestein}, {Forin}, {Gatlin},
  {Ghandi}, {Heil}, {Holohan}, {El Husseini}, {Juhasz}, {Kagi}, {Kovvuri},
  {Lanka}, {van Megen}, {Mukhortov}, {Patel}, {Perez}, {Rapsang}, {Reinhardt},
  {Rouhani}, {Sapek}, {Seera}, {Shekar}, {Sridharan}, {Weisz}, {Woods}, {Yi
  Xiao}, {Zhang}, {Zhao}, and {Burger}}{{Chung} et~al\mbox{.}}{2018}]%
        {brainwave}
\bibfield{author}{\bibinfo{person}{E. {Chung}}, \bibinfo{person}{J. {Fowers}},
  \bibinfo{person}{K. {Ovtcharov}}, \bibinfo{person}{M. {Papamichael}},
  \bibinfo{person}{A. {Caulfield}}, \bibinfo{person}{T. {Massengill}},
  \bibinfo{person}{M. {Liu}}, \bibinfo{person}{D. {Lo}}, \bibinfo{person}{S.
  {Alkalay}}, \bibinfo{person}{M. {Haselman}}, \bibinfo{person}{M.
  {Abeydeera}}, \bibinfo{person}{L. {Adams}}, \bibinfo{person}{H. {Angepat}},
  \bibinfo{person}{C. {Boehn}}, \bibinfo{person}{D. {Chiou}},
  \bibinfo{person}{O. {Firestein}}, \bibinfo{person}{A. {Forin}},
  \bibinfo{person}{K.~S. {Gatlin}}, \bibinfo{person}{M. {Ghandi}},
  \bibinfo{person}{S. {Heil}}, \bibinfo{person}{K. {Holohan}},
  \bibinfo{person}{A. {El Husseini}}, \bibinfo{person}{T. {Juhasz}},
  \bibinfo{person}{K. {Kagi}}, \bibinfo{person}{R. {Kovvuri}},
  \bibinfo{person}{S. {Lanka}}, \bibinfo{person}{F. {van Megen}},
  \bibinfo{person}{D. {Mukhortov}}, \bibinfo{person}{P. {Patel}},
  \bibinfo{person}{B. {Perez}}, \bibinfo{person}{A. {Rapsang}},
  \bibinfo{person}{S. {Reinhardt}}, \bibinfo{person}{B. {Rouhani}},
  \bibinfo{person}{A. {Sapek}}, \bibinfo{person}{R. {Seera}},
  \bibinfo{person}{S. {Shekar}}, \bibinfo{person}{B. {Sridharan}},
  \bibinfo{person}{G. {Weisz}}, \bibinfo{person}{L. {Woods}},
  \bibinfo{person}{P. {Yi Xiao}}, \bibinfo{person}{D. {Zhang}},
  \bibinfo{person}{R. {Zhao}}, {and} \bibinfo{person}{D. {Burger}}.}
  \bibinfo{year}{2018}\natexlab{}.
\newblock \showarticletitle{Serving DNNs in Real Time at Datacenter Scale with
  Project Brainwave}.
\newblock \bibinfo{journal}{\emph{IEEE Micro}} \bibinfo{volume}{38},
  \bibinfo{number}{2} (\bibinfo{date}{Mar} \bibinfo{year}{2018}),
  \bibinfo{pages}{8--20}.
\newblock
\showISSN{0272-1732}
\urldef\tempurl%
\url{https://doi.org/10.1109/MM.2018.022071131}
\showDOI{\tempurl}


\bibitem[\protect\citeauthoryear{{Cong}, {Liu}, {Neuendorffer}, {Noguera},
  {Vissers}, and {Zhang}}{{Cong} et~al\mbox{.}}{2011}]%
        {xilinx_hls}
\bibfield{author}{\bibinfo{person}{J. {Cong}}, \bibinfo{person}{B. {Liu}},
  \bibinfo{person}{S. {Neuendorffer}}, \bibinfo{person}{J. {Noguera}},
  \bibinfo{person}{K. {Vissers}}, {and} \bibinfo{person}{Z. {Zhang}}.}
  \bibinfo{year}{2011}\natexlab{}.
\newblock \showarticletitle{High-Level Synthesis for {FPGAs}: From Prototyping
  to Deployment}.
\newblock \bibinfo{journal}{\emph{IEEE Transactions on Computer-Aided Design of
  Integrated Circuits and Systems}} \bibinfo{volume}{30}, \bibinfo{number}{4}
  (\bibinfo{date}{April} \bibinfo{year}{2011}), \bibinfo{pages}{473--491}.
\newblock
\showISSN{0278-0070}
\urldef\tempurl%
\url{https://doi.org/10.1109/TCAD.2011.2110592}
\showDOI{\tempurl}


\bibitem[\protect\citeauthoryear{Czajkowski, Aydonat, Denisenko, Freeman,
  Kinsner, Neto, Wong, Yiannacouras, and Singh}{Czajkowski
  et~al\mbox{.}}{2012}]%
        {intel_opencl}
\bibfield{author}{\bibinfo{person}{Tomasz~S Czajkowski}, \bibinfo{person}{Utku
  Aydonat}, \bibinfo{person}{Dmitry Denisenko}, \bibinfo{person}{John Freeman},
  \bibinfo{person}{Michael Kinsner}, \bibinfo{person}{David Neto},
  \bibinfo{person}{Jason Wong}, \bibinfo{person}{Peter Yiannacouras}, {and}
  \bibinfo{person}{Deshanand~P Singh}.} \bibinfo{year}{2012}\natexlab{}.
\newblock \showarticletitle{From {OpenCL} to high-performance hardware on
  {FPGAs}}. In \bibinfo{booktitle}{\emph{22nd international conference on field
  programmable logic and applications (FPL)}}. IEEE, \bibinfo{pages}{531--534}.
\newblock


\bibitem[\protect\citeauthoryear{de~Fine~Licht, Meierhans, and
  Hoefler}{de~Fine~Licht et~al\mbox{.}}{2018}]%
        {hls_transformations}
\bibfield{author}{\bibinfo{person}{Johannes de Fine~Licht},
  \bibinfo{person}{Simon Meierhans}, {and} \bibinfo{person}{Torsten Hoefler}.}
  \bibinfo{year}{2018}\natexlab{}.
\newblock \showarticletitle{Transformations of High-Level Synthesis Codes for
  High-Performance Computing}.
\newblock \bibinfo{journal}{\emph{CoRR}}  \bibinfo{volume}{abs/1805.08288}
  (\bibinfo{year}{2018}).
\newblock
\showeprint[arxiv]{1805.08288}
\urldef\tempurl%
\url{http://arxiv.org/abs/1805.08288}
\showURL{%
\tempurl}


\bibitem[\protect\citeauthoryear{Dimond, Racaniere, and Pell}{Dimond
  et~al\mbox{.}}{2011}]%
        {maxeler}
\bibfield{author}{\bibinfo{person}{Rob Dimond}, \bibinfo{person}{S{\'e}bastien
  Racaniere}, {and} \bibinfo{person}{Oliver Pell}.}
  \bibinfo{year}{2011}\natexlab{}.
\newblock \showarticletitle{Accelerating large-scale HPC Applications using
  FPGAs}. In \bibinfo{booktitle}{\emph{2011 IEEE 20th Symposium on Computer
  Arithmetic}}. IEEE, \bibinfo{pages}{191--192}.
\newblock


\bibitem[\protect\citeauthoryear{{Domke}, {Hoefler}, and {Nagel}}{{Domke}
  et~al\mbox{.}}{2011}]%
        {deadlock-free}
\bibfield{author}{\bibinfo{person}{J. {Domke}}, \bibinfo{person}{T. {Hoefler}},
  {and} \bibinfo{person}{W.~E. {Nagel}}.} \bibinfo{year}{2011}\natexlab{}.
\newblock \showarticletitle{Deadlock-Free Oblivious Routing for Arbitrary
  Topologies}. In \bibinfo{booktitle}{\emph{2011 IEEE International Parallel
  Distributed Processing Symposium}}. \bibinfo{pages}{616--627}.
\newblock
\showISSN{1530-2075}
\urldef\tempurl%
\url{https://doi.org/10.1109/IPDPS.2011.65}
\showDOI{\tempurl}


\bibitem[\protect\citeauthoryear{Eskandari, Tarafdar, Ly-Ma, and
  Chow}{Eskandari et~al\mbox{.}}{2019}]%
        {galapagos}
\bibfield{author}{\bibinfo{person}{Nariman Eskandari}, \bibinfo{person}{Naif
  Tarafdar}, \bibinfo{person}{Daniel Ly-Ma}, {and} \bibinfo{person}{Paul
  Chow}.} \bibinfo{year}{2019}\natexlab{}.
\newblock \showarticletitle{A Modular Heterogeneous Stack for Deploying FPGAs
  and CPUs in the Data Center}. In \bibinfo{booktitle}{\emph{Proceedings of the
  2019 ACM/SIGDA International Symposium on Field-Programmable Gate Arrays}}
  \emph{(\bibinfo{series}{FPGA '19})}. \bibinfo{publisher}{ACM},
  \bibinfo{address}{New York, NY, USA}, \bibinfo{pages}{262--271}.
\newblock
\showISBNx{978-1-4503-6137-8}
\urldef\tempurl%
\url{https://doi.org/10.1145/3289602.3293909}
\showDOI{\tempurl}


\bibitem[\protect\citeauthoryear{{Geng}, {Wang}, {Sanaullah}, {Yang}, {Xu},
  {Patel}, and {Herbordt}}{{Geng} et~al\mbox{.}}{2018}]%
        {fdeep}
\bibfield{author}{\bibinfo{person}{T. {Geng}}, \bibinfo{person}{T. {Wang}},
  \bibinfo{person}{A. {Sanaullah}}, \bibinfo{person}{C. {Yang}},
  \bibinfo{person}{R. {Xu}}, \bibinfo{person}{R. {Patel}}, {and}
  \bibinfo{person}{M. {Herbordt}}.} \bibinfo{year}{2018}\natexlab{}.
\newblock \showarticletitle{FPDeep: Acceleration and Load Balancing of CNN
  Training on FPGA Clusters}. In \bibinfo{booktitle}{\emph{2018 IEEE 26th
  Annual International Symposium on Field-Programmable Custom Computing
  Machines (FCCM)}}. \bibinfo{pages}{81--84}.
\newblock
\showISSN{2576-2621}


\bibitem[\protect\citeauthoryear{{George}, {Herbordt}, {Lam}, {Lawande},
  {Sheng}, and {Yang}}{{George} et~al\mbox{.}}{2016}]%
        {novo-g-sharp}
\bibfield{author}{\bibinfo{person}{A.~D. {George}}, \bibinfo{person}{M.~C.
  {Herbordt}}, \bibinfo{person}{H. {Lam}}, \bibinfo{person}{A.~G. {Lawande}},
  \bibinfo{person}{J. {Sheng}}, {and} \bibinfo{person}{C. {Yang}}.}
  \bibinfo{year}{2016}\natexlab{}.
\newblock \showarticletitle{Novo-G\#: Large-scale reconfigurable computing with
  direct and programmable interconnects}. In \bibinfo{booktitle}{\emph{2016
  IEEE High Performance Extreme Computing Conference (HPEC)}}.
  \bibinfo{pages}{1--7}.
\newblock


\bibitem[\protect\citeauthoryear{{Gysi}, {Bär}, and {Hoefler}}{{Gysi}
  et~al\mbox{.}}{2016}]%
        {dcuda}
\bibfield{author}{\bibinfo{person}{T. {Gysi}}, \bibinfo{person}{J. {Bär}},
  {and} \bibinfo{person}{T. {Hoefler}}.} \bibinfo{year}{2016}\natexlab{}.
\newblock \showarticletitle{dCUDA: Hardware Supported Overlap of Computation
  and Communication}. In \bibinfo{booktitle}{\emph{SC '16: Proceedings of the
  International Conference for High Performance Computing, Networking, Storage
  and Analysis}}. \bibinfo{pages}{609--620}.
\newblock
\showISSN{2167-4337}
\urldef\tempurl%
\url{https://doi.org/10.1109/SC.2016.51}
\showDOI{\tempurl}


\bibitem[\protect\citeauthoryear{instances}{instances}{[n. d.]}]%
        {amazon_f1}
\bibfield{author}{\bibinfo{person}{Amazon EC2~F1 instances}.}
  \bibinfo{year}{[n. d.]}\natexlab{}.
\newblock
  \bibinfo{howpublished}{\url{https://aws.amazon.com/ec2/instance-types/f1/}}.
\newblock


\bibitem[\protect\citeauthoryear{Kara, Alistarh, Alonso, Mutlu, and Zhang}{Kara
  et~al\mbox{.}}{2017}]%
        {kara2017fpga}
\bibfield{author}{\bibinfo{person}{Kaan Kara}, \bibinfo{person}{Dan Alistarh},
  \bibinfo{person}{Gustavo Alonso}, \bibinfo{person}{Onur Mutlu}, {and}
  \bibinfo{person}{Ce Zhang}.} \bibinfo{year}{2017}\natexlab{}.
\newblock \showarticletitle{FPGA-accelerated dense linear machine learning: A
  precision-convergence trade-off}. In \bibinfo{booktitle}{\emph{2017 IEEE 25th
  Annual International Symposium on Field-Programmable Custom Computing
  Machines (FCCM)}}. IEEE, \bibinfo{pages}{160--167}.
\newblock


\bibitem[\protect\citeauthoryear{Kim, Lee, and Vetter}{Kim
  et~al\mbox{.}}{2016}]%
        {impacc}
\bibfield{author}{\bibinfo{person}{Jungwon Kim}, \bibinfo{person}{Seyong Lee},
  {and} \bibinfo{person}{Jeffrey~S. Vetter}.} \bibinfo{year}{2016}\natexlab{}.
\newblock \showarticletitle{IMPACC: A Tightly Integrated MPI+OpenACC Framework
  Exploiting Shared Memory Parallelism}. In
  \bibinfo{booktitle}{\emph{Proceedings of the 25th ACM International Symposium
  on High-Performance Parallel and Distributed Computing}}
  \emph{(\bibinfo{series}{HPDC '16})}. \bibinfo{publisher}{ACM},
  \bibinfo{address}{New York, NY, USA}, \bibinfo{pages}{189--201}.
\newblock
\showISBNx{978-1-4503-4314-5}
\urldef\tempurl%
\url{https://doi.org/10.1145/2907294.2907302}
\showDOI{\tempurl}


\bibitem[\protect\citeauthoryear{Kobayashi, Oobata, Fujita, Yamaguchi, and
  Boku}{Kobayashi et~al\mbox{.}}{2018}]%
        {opecl_ready_network}
\bibfield{author}{\bibinfo{person}{Ryohei Kobayashi}, \bibinfo{person}{Yuma
  Oobata}, \bibinfo{person}{Norihisa Fujita}, \bibinfo{person}{Yoshiki
  Yamaguchi}, {and} \bibinfo{person}{Taisuke Boku}.}
  \bibinfo{year}{2018}\natexlab{}.
\newblock \showarticletitle{OpenCL-ready High Speed FPGA Network for
  Reconfigurable High Performance Computing}. In
  \bibinfo{booktitle}{\emph{Proceedings of the International Conference on High
  Performance Computing in Asia-Pacific Region}} \emph{(\bibinfo{series}{HPC
  Asia 2018})}. \bibinfo{publisher}{ACM}, \bibinfo{address}{New York, NY, USA},
  \bibinfo{pages}{192--201}.
\newblock
\showISBNx{978-1-4503-5372-4}
\urldef\tempurl%
\url{https://doi.org/10.1145/3149457.3149479}
\showDOI{\tempurl}


\bibitem[\protect\citeauthoryear{{Lawande}, {George}, and {Lam}}{{Lawande}
  et~al\mbox{.}}{2016}]%
        {novo-g-sharp-opencl}
\bibfield{author}{\bibinfo{person}{A. {Lawande}}, \bibinfo{person}{A.~D.
  {George}}, {and} \bibinfo{person}{H. {Lam}}.}
  \bibinfo{year}{2016}\natexlab{}.
\newblock \showarticletitle{An OpenCL Framework for Distributed Apps on a
  Multidimensional Network of FPGAs}. In \bibinfo{booktitle}{\emph{2016 6th
  Workshop on Irregular Applications: Architecture and Algorithms (IA3)}}.
  \bibinfo{pages}{42--49}.
\newblock
\urldef\tempurl%
\url{https://doi.org/10.1109/IA3.2016.012}
\showDOI{\tempurl}


\bibitem[\protect\citeauthoryear{Matteis, de~Fine~Licht, and Hoefler}{Matteis
  et~al\mbox{.}}{2019}]%
        {fblas}
\bibfield{author}{\bibinfo{person}{Tiziano~De Matteis},
  \bibinfo{person}{Johannes de Fine~Licht}, {and} \bibinfo{person}{Torsten
  Hoefler}.} \bibinfo{year}{2019}\natexlab{}.
\newblock \showarticletitle{FBLAS: Streaming Linear Algebra on FPGA}.
\newblock \bibinfo{journal}{\emph{CoRR}} (\bibinfo{date}{Aug.}
  \bibinfo{year}{2019}).
\newblock


\bibitem[\protect\citeauthoryear{{Message Passing Interface Forum}}{{Message
  Passing Interface Forum}}{2015}]%
        {mpi_standard}
\bibfield{author}{\bibinfo{person}{{Message Passing Interface Forum}}.}
  \bibinfo{year}{2015}\natexlab{}.
\newblock \bibinfo{booktitle}{\emph{MPI: A Message-Passing Interface Standard,
  Version 3.1}}.
\newblock \bibinfo{type}{Specification}.
\newblock
\urldef\tempurl%
\url{https://www.mpi-forum.org/docs/mpi-3.1/mpi31-report.pdf}
\showURL{%
\tempurl}


\bibitem[\protect\citeauthoryear{{Owaida} and {Alonso}}{{Owaida} and
  {Alonso}}{2018}]%
        {inference_tree}
\bibfield{author}{\bibinfo{person}{M. {Owaida}} {and} \bibinfo{person}{G.
  {Alonso}}.} \bibinfo{year}{2018}\natexlab{}.
\newblock \showarticletitle{Application Partitioning on FPGA Clusters:
  Inference over Decision Tree Ensembles}. In \bibinfo{booktitle}{\emph{2018
  28th International Conference on Field Programmable Logic and Applications
  (FPL)}}. \bibinfo{pages}{295--2955}.
\newblock
\showISSN{1946-1488}
\urldef\tempurl%
\url{https://doi.org/10.1109/FPL.2018.00057}
\showDOI{\tempurl}


\bibitem[\protect\citeauthoryear{Salda\~{n}a, Patel, Madill, Nunes, Wang, Chow,
  Wittig, Styles, and Putnam}{Salda\~{n}a et~al\mbox{.}}{2010}]%
        {tmd-mpi}
\bibfield{author}{\bibinfo{person}{Manuel Salda\~{n}a}, \bibinfo{person}{Arun
  Patel}, \bibinfo{person}{Christopher Madill}, \bibinfo{person}{Daniel Nunes},
  \bibinfo{person}{Danyao Wang}, \bibinfo{person}{Paul Chow},
  \bibinfo{person}{Ralph Wittig}, \bibinfo{person}{Henry Styles}, {and}
  \bibinfo{person}{Andrew Putnam}.} \bibinfo{year}{2010}\natexlab{}.
\newblock \showarticletitle{MPI As a Programming Model for High-Performance
  Reconfigurable Computers}.
\newblock \bibinfo{journal}{\emph{ACM Trans. Reconfigurable Technol. Syst.}}
  \bibinfo{volume}{3}, \bibinfo{number}{4}, Article \bibinfo{articleno}{22}
  (\bibinfo{date}{Nov.} \bibinfo{year}{2010}), \bibinfo{numpages}{29}~pages.
\newblock
\showISSN{1936-7406}
\urldef\tempurl%
\url{https://doi.org/10.1145/1862648.1862652}
\showDOI{\tempurl}


\bibitem[\protect\citeauthoryear{Sano, Hatsuda, and Yamamoto}{Sano
  et~al\mbox{.}}{2014}]%
        {sano2014multi}
\bibfield{author}{\bibinfo{person}{Kentaro Sano}, \bibinfo{person}{Yoshiaki
  Hatsuda}, {and} \bibinfo{person}{Satoru Yamamoto}.}
  \bibinfo{year}{2014}\natexlab{}.
\newblock \showarticletitle{Multi-FPGA accelerator for scalable stencil
  computation with constant memory bandwidth}.
\newblock \bibinfo{journal}{\emph{IEEE Transactions on Parallel and Distributed
  Systems}} \bibinfo{volume}{25}, \bibinfo{number}{3} (\bibinfo{year}{2014}),
  \bibinfo{pages}{695--705}.
\newblock


\bibitem[\protect\citeauthoryear{Shu, Cheng, Chen, Guo, Qu, Xiong, Chiou, and
  Moscibroda}{Shu et~al\mbox{.}}{2019}]%
        {dua}
\bibfield{author}{\bibinfo{person}{Ran Shu}, \bibinfo{person}{Peng Cheng},
  \bibinfo{person}{Guo Chen}, \bibinfo{person}{Zhiyuan Guo},
  \bibinfo{person}{Lei Qu}, \bibinfo{person}{Yongqiang Xiong},
  \bibinfo{person}{Derek Chiou}, {and} \bibinfo{person}{Thomas Moscibroda}.}
  \bibinfo{year}{2019}\natexlab{}.
\newblock \showarticletitle{Direct Universal Access: Making Data Center
  Resources Available to {FPGA}}. In \bibinfo{booktitle}{\emph{16th {USENIX}
  Symposium on Networked Systems Design and Implementation ({NSDI} 19)}}.
  \bibinfo{publisher}{{USENIX} Association}, \bibinfo{address}{Boston, MA},
  \bibinfo{pages}{127--140}.
\newblock
\showISBNx{978-1-931971-49-2}
\urldef\tempurl%
\url{https://www.usenix.org/conference/nsdi19/presentation/shu}
\showURL{%
\tempurl}


\bibitem[\protect\citeauthoryear{Table}{Table}{[n. d.]a}]%
        {stratix10_product_table}
\bibfield{author}{\bibinfo{person}{Stratix 10 GX/SX~Product Table}.}
  \bibinfo{year}{[n. d.]}\natexlab{a}.
\newblock
  \bibinfo{howpublished}{\url{https://www.intel.com/content/dam/www/programmable/us/en/pdfs/literature/pt/stratix-10-product-table.pdf}}.
\newblock


\bibitem[\protect\citeauthoryear{Table}{Table}{[n. d.]b}]%
        {versal_product_table}
\bibfield{author}{\bibinfo{person}{Versal ACAP AI Core Series~Product Table}.}
  \bibinfo{year}{[n. d.]}\natexlab{b}.
\newblock
  \bibinfo{howpublished}{\url{https://www.xilinx.com/support/documentation/selection-guides/versal-ai-core-product-selection-guide.pdf}}.
\newblock


\bibitem[\protect\citeauthoryear{Umuroglu, Fraser, Gambardella, Blott, Leong,
  Jahre, and Vissers}{Umuroglu et~al\mbox{.}}{2017}]%
        {finn}
\bibfield{author}{\bibinfo{person}{Yaman Umuroglu},
  \bibinfo{person}{Nicholas~J. Fraser}, \bibinfo{person}{Giulio Gambardella},
  \bibinfo{person}{Michaela Blott}, \bibinfo{person}{Philip Leong},
  \bibinfo{person}{Magnus Jahre}, {and} \bibinfo{person}{Kees Vissers}.}
  \bibinfo{year}{2017}\natexlab{}.
\newblock \showarticletitle{FINN: A Framework for Fast, Scalable Binarized
  Neural Network Inference}. In \bibinfo{booktitle}{\emph{Proceedings of the
  2017 ACM/SIGDA International Symposium on Field-Programmable Gate Arrays}}
  \emph{(\bibinfo{series}{FPGA '17})}. \bibinfo{publisher}{ACM},
  \bibinfo{address}{New York, NY, USA}, \bibinfo{pages}{65--74}.
\newblock
\showISBNx{978-1-4503-4354-1}
\urldef\tempurl%
\url{https://doi.org/10.1145/3020078.3021744}
\showDOI{\tempurl}


\bibitem[\protect\citeauthoryear{Zhang, Wu, Sun, Sun, Luo, and Cong}{Zhang
  et~al\mbox{.}}{2016}]%
        {cnn_cluster_fpga}
\bibfield{author}{\bibinfo{person}{Chen Zhang}, \bibinfo{person}{Di Wu},
  \bibinfo{person}{Jiayu Sun}, \bibinfo{person}{Guangyu Sun},
  \bibinfo{person}{Guojie Luo}, {and} \bibinfo{person}{Jason Cong}.}
  \bibinfo{year}{2016}\natexlab{}.
\newblock \showarticletitle{Energy-Efficient CNN Implementation on a Deeply
  Pipelined FPGA Cluster}. In \bibinfo{booktitle}{\emph{Proceedings of the 2016
  International Symposium on Low Power Electronics and Design}}
  \emph{(\bibinfo{series}{ISLPED '16})}. \bibinfo{publisher}{ACM},
  \bibinfo{address}{New York, NY, USA}, \bibinfo{pages}{326--331}.
\newblock
\showISBNx{978-1-4503-4185-1}
\urldef\tempurl%
\url{https://doi.org/10.1145/2934583.2934644}
\showDOI{\tempurl}


\bibitem[\protect\citeauthoryear{Zohouri, Podobas, and Matsuoka}{Zohouri
  et~al\mbox{.}}{2018}]%
        {zohouri2018combined}
\bibfield{author}{\bibinfo{person}{Hamid~Reza Zohouri}, \bibinfo{person}{Artur
  Podobas}, {and} \bibinfo{person}{Satoshi Matsuoka}.}
  \bibinfo{year}{2018}\natexlab{}.
\newblock \showarticletitle{Combined spatial and temporal blocking for
  high-performance stencil computation on FPGAs using OpenCL}. In
  \bibinfo{booktitle}{\emph{Proceedings of the 2018 ACM/SIGDA International
  Symposium on Field-Programmable Gate Arrays}}. ACM,
  \bibinfo{pages}{153--162}.
\newblock


\end{thebibliography}
  
\end{document}